\documentclass[manuscript]{aastex}

\usepackage{natbib, aas_macros}
\usepackage{amsmath}
\usepackage[usenames]{color}
\usepackage{ulem}

\definecolor{MyDarkBlue}{rgb}{0,0.08,0.45}
\definecolor{MyDarkRed}{rgb}{0.8,0.1,0.08}
\definecolor{Red}{rgb}{1.0,0.0,0.2}
\definecolor{Blue}{rgb}{0,0.08,0.95}
\definecolor{LightGrey}{rgb}{0.7,0.7,0.7}

\begin{document}
\title{High-resolution calculations of the solar global convection with
the reduced speed of sound technique: I. The structure of the convection and
the magnetic field without the rotation.}
\author{H. Hotta$^1$, M. Rempel$^2$ and T. Yokoyama$^1$}
\affil{$^1$Department of Earth and Planetary Science, University of Tokyo,
7-3-1 Hongo, Bunkyo-ku, Tokyo 113-0033, Japan\\
$^2$High Altitude Observatory, National Center for Atmospheric Research,
Boulder, CO, USA}
\email{ hotta.h@eps.s.u-tokyo.ac.jp}
\begin{abstract}
We carry out non-rotating high-resolution calculations of the solar global
convection, which resolve convective scales of less than 10 Mm. To cope
with the low Mach number conditions in the lower convection zone, we use the
reduced speed of sound technique (RSST), which is simple to implement
and requires only local communication in the parallel computation.
In addition, the RSST allows us to expand the computational domain
upward to about $0.99 R_{\odot}$ as it can also handle compressible
flows.
Using this approach, we study the solar convection zone on the global scale, 
including 
small-scale near-surface convection. In particular, we investigate the influence of the top
boundary condition on the convective structure throughout the
convection zone as well as on small-scale dynamo action.
Our main conclusions are: 1. The small-scale downflows generated in the
 near-surface layer penetrate into deeper layers to some extent
and excite small-scale turbulence
in the region $>0.9R_\odot$, where $R_\odot$ is the solar radius.
 2. In the deeper convection
zone ($<0.9R_\odot$), the convection is not influenced by the
location of the upper boundary.
3. Using an LES approach, we can achieve small-scale dynamo action and
maintain a field of about $0.15-0.25 B_\mathrm{eq}$ throughout the convection
zone, where $B_\mathrm{eq}$ is the equipartition magnetic
 field to the kinetic energy.
4. The overall dynamo efficiency varies significantly in the convection
zone as a consequence of the downward directed Poynting flux and the
depth variation of the intrinsic convective scales.
\end{abstract}
\keywords{Sun: interior --- Sun: dynamo --- Stars: interiors}
\section{INTRODUCTION}
The solar convection zone is filled with turbulent thermal
convection due to its superadiabatic stratification, which is
maintained by the radiative energy flux from the core of the Sun.
Due to the influence of rotation,
the anisotropic convection transports angular momentum as well
as energy. This maintains mean flows such as the meridional flow
and the differential rotation. These flows are thought to be
important for the dynamo and related eleven-year cyclic variations of the
sunspot numbers
\citep{1955ApJ...122..293P,1995A&A...303L..29C,1999ApJ...518..508D,
2010ApJ...709.1009H,2010ApJ...714L.308H}. 
In addition, the turbulent thermal convection itself is also important
for the generation of the magnetic field.
Idealized simulations
\citep[e.g.,][]{1996JFM...306..325B,1999ApJ...515L..39C} and the realistic
calculations for the photosphere
\citep{2007A&A...465L..43V,0004-637X-714-2-1606} have suggested that
chaotic motions
have the ability to maintain magnetic energy through a small-scale dynamo process. Such
dynamos have not been studied so far on the the global scale, i.e., the entire convection zone.
\par
The purpose of this study is to achieve
multi-scale convection, including scales from $10$~Mm (smaller than 
supergranulation) up to $\sim 200$~Mm (global scale). We accomplish this by approaching the solar surface
up to $0.99R_{\odot}$ using unprecedented resolution, studying in particular the influence of the location 
of the upper boundary on the convective structure in the deeper parts of the convection zone.
In addition, we use
our approach to study the transport and the generation of  the magnetic field by turbulent
convection in the absence of rotation, i.e., we study the operation of a small-scale dynamo
in the highly stratified convection zone.
\par
The paper is organized as follows:
We provide an overview of the reduced speed of sound technique in \S \ref{S:rsst}, 
introduce our numerical setting in \S \ref{model}, and show the results
in \S \ref{result},present the properties based on the obtained spatial distributions in \S \ref{structure}, discuss the energy balance
using the integrated flux in \S \ref{integrate_flux}, analyze
, the properties of the convection in cases without the magnetic fields
using the spherical harmonic expansion in \S \ref{sh}, and analyze
the cases with magnetic fields
using spherical harmonic expansion and the probability density function
in \S \ref{sh_mag}, investigate 
the transportation and the generation of the magnetic field in the
convection zone in \S \ref{local_dynamo}, and 
 summarize our investigation in \S \ref{discussion}.
\section{Reduced speed of sound technique}\label{S:rsst}
Developing numerical simulations of solar global convection
is challenging due to the high resolution and the broad range
of dynamical time scales that are required.
One of the most paramount 
difficulties arises from the high speed of sound and the related low Mach-number 
flows throughout most of the convection zone.
At the base of the convection zone, the speed of sound
is about $200\ \mathrm{km\ s^{-1}}$, while the speed of convection
is thought to be $50\ \mathrm{m\ s^{-1}}$
\citep[e.g.,][]{2004suin.book.....S}. The time step must therefore be shorter
owing to the CFL (Courant-Friedrichs-Lewy) condition in an explicit fully compressible method,
even when we are interested in the phenomena related to convection. In
order to avoid this situation, the anelastic approximation
is frequently adopted in which the mass conservation equation is
replaced with $\nabla\cdot (\rho_0{\bf v})=0$, where $\rho_0$
is the reference density and ${\bf v}$ is the fluid velocity. In this
approximation, the speed of sound is assumed to be infinite, and
we need to solve the elliptic equation for the pressure, which
filters out the propagation of the sound wave. Since the anelastic
approximation is well applicable deep in the convection zone and the
time step is no longer limited by the large speed of sound, the solar
global convection has been investigated with this method in numerous 
studies about the differential rotation, the meridional flow, the global
dynamo, and the dynamical coupling of the radiative zone
\citep{2000ApJ...532..593M,2006ApJ...641..618M,2008ApJ...673..557M,2002ApJ...570..865B,
2004ApJ...614.1073B,2011ApJ...742...79B,2006ApJ...648L.157B,2010ApJ...715L.133G}.\par
There are, however, two drawbacks in the anelastic approximation. The
first is the breakdown of the approximation near the solar surface. Since
the convection velocity becomes higher and the speed of sound becomes
lower in the near-surface layer ($>0.98R_\odot$), these have similar values and the anelastic
approximation cannot be applied. Since the interaction
between the small-scale convection generated in the near-surface layer
(granulation and supergranulation) and large scale convection  (giant
cell and global one) is an important topic, the
connection between  the near-surface layer and the global convection has been investigated
\citep[e.g.,][]{2011JPhCS.271a2070A}. The global calculation, however,
which includes all such multiple scales, has not been achieved
yet. The second
drawback is the difficulty arising from the increase of the required resolution.
The pseudo-spectral method based on the spherical harmonic
expansion is frequently adopted, especially for solving the elliptic
equation of the pressure. In this method, non-linear terms require the
transformation of the physical variable from the real space to the
spectral space, and vice versa every time step. The calculation cost of the
transformation is estimated as 
$\mathcal{O}(N_\theta^2N_\phi\log N_\phi)$ due to the absence of a 
fast algorithm for the Legendre transformation, which is as powerful as the
fast Fourier transformation (FFT), where $N_\theta$ and $N_\phi$
are the maximum mode numbers in the latitude and the longitude respectively.
Thus the computational cost of this method 
is significant, limiting  the achievable
resolution.
Owing 
to this, some numerical calculations of the
geodynamo adopt the finite difference method in order to achieve
high resolution \citep{2008Natur.454.1106K,2010Natur.463..793M}.
As explained above, when the near-surface layer is included in the calculation,
the typical convection scale becomes small and a large number of
grid points is required. The resolution is one of the keys to
access the solar surface.
We note that some studies using the finite difference have been done in the stellar
or solar context using a
moderate ratio of the speed of sound and the convection velocity
with adjustments of the
radiative flux and the stratification
\citep{2011A&A...531A.162K,2012ApJ...755L..22K}.
Although this type of
approach provides insights for the maintenance of the mean flow and the
magnetic field, proper reproduction using the solar parameters, such
as the stratification, the luminosity, the partial
ionization effect and the
rotation, and direct comparison with actual observations
 cannot be
achieved.\par
The reduced speed of sound technique
\cite[hereafter
RSST,][]{2005ApJ...622.1320R,2006ApJ...647..662R,2012A&A...539A..30H}
can overcome
these drawbacks while avoiding the severe time step caused by the speed
of sound. In the RSST the equation of continuity is replaced with
\begin{eqnarray}
 \frac{\partial \rho}{\partial t}=-\frac{1}{\xi^2}\nabla\cdot(\rho {\bf v}).
\end{eqnarray}
The speed of sound is then reduced $\xi$ times, but the dispersion
relationship for sound waves remains (where wave speed
decreases equally for all wavelengths). This technique does not change the
hyperbolic character of the equations that can be integrated
explicitly. Due to this hyperbolicity, only the
local communication 
is required, which decreases the communication overhead
in parallel computing. A simple algorithm and low cost of
communication are significant factors
that enable high resolution calculations. \cite{2012A&A...539A..30H} investigated the validity of the
RSST in a thermal convection problem. It was concluded that the RSST is
valid when the Mach number defined with the RMS (root mean square)
velocity and the reduced speed of sound $\hat{c_s}$ is smaller than 0.7.
Another advantage of this method is its accessibility to the real solar
surface with an inhomogeneous $\xi$. 
The Mach number varies
substantially in the solar convection zone. When a moderate or zero 
reduction of the speed of sound is used in the near-surface layer while
a large $\xi$ around the bottom part of the convection zone is used, the
properties of the thermal convection, even including the surface, can be properly investigated
without undermining the physics.
\cite{2012A&A...539A..30H} confirmed that the inhomogeneous $\xi$ is valid
again when the Mach number is less than 0.7.
\section{MODEL}\label{model}
We solve the three-dimensional magnetohydrodynamic equations with the
RSST in the spherical geometry $(r,\theta,\phi)$ as:
\begin{eqnarray}
&&\frac{\partial \rho_1}{\partial t} =
 -\frac{1}{\xi^2}\nabla\cdot\left(\rho_0 {\bf
			     v}\right),\label{eqco}\\
&&\rho_0\frac{\partial {\bf v}}{\partial t} = -\rho_0({\bf v}\cdot\nabla){\bf v} -
 \nabla\left( p_1 + \frac{B^2}{8\pi}\right)
 +\nabla\cdot\left(\frac{{\bf BB}}{4\pi}\right)
 - \rho_1g{\bf e_r},\label{eqmo}\\
&& \frac{\partial {\bf B}}{\partial t}=\nabla\times\left({\bf v}\times
						  {\bf B}\right),\label{eqma}\\
&& \rho_0 T_0\frac{\partial s_1}{\partial t} = -\rho_0 T_0({\bf v}\cdot\nabla)s_1
 +
 \frac{1}{r^2}
\frac{d}{dr}
\left(r^2
  \kappa_\mathrm{r}\rho_0 c_\mathrm{p}\frac{dT_0}{dr} \right) + \Gamma,\label{eqen}
\end{eqnarray}
where $\rho$, $p$, $s$, $T$, ${\bf v}$, and ${\bf B}$ are the
density, the gas pressure, the specific entropy, the temperature, the fluid velocity and
the magnetic field, respectively. The subscript 1 denotes the
fluctuation from the time-independent spherically symmetric reference
state, which has subscript 0. 
$g$, $\kappa_\mathrm{r}$, and $\Gamma$ are
the gravitational acceleration, the coefficient of the radiative
diffusivity, and the surface cooling term, respectively. The setting of
these values is explained in the following
paragraph.
In order to close the MHD system, we adopt a non-ideal equation of state as:
\begin{eqnarray}
 p_1 = \left(\frac{\partial p}{\partial \rho}\right)_s   \rho_1 
       + \left(\frac{\partial p}{\partial s   }\right)_\rho s_1.
\end{eqnarray}
Since our upper boundary is at $r=0.99R_\odot$ at maximum, in the near-surface
region
partial ionization is
important (see Fig. \ref{back}f)
and is included in our treatment (see Appendix \ref{app:eos} for details) by using the OPAL
repository with the solar abundance of $X=0.75$, $Y=0.23$, and $Z=0.02$,
where $X$, $Y$, and $Z$ specify the mass fraction of hydrogen, helium
and other metals, respectively.\par
Note that we do not have any explicit turbulent
thermal diffusivity and viscosity \citep{2000ApJ...532..593M} in order
to maximize the fluid and
magnetic Reynolds number, but use the artificial viscosity introduced
in \cite{2009ApJ...691..640R}. 
Our treatment ensures
that the dissipated kinetic and magnetic energies with the artificial
viscosity are converted to the internal energy. Note also that
the rotation is not included in this study, since we focus on the
connection of thermal convection between the small and large scales and
its local dynamo action on the global scale. 
The combination of
rotation and stratified convection leads to differential rotation and additional 
turbulent induction effects that enable large-scale dynamo action. It is difficult to
distinguish the local dynamo action
from global dynamo action if rotation is
present.
Since the rotation effect in the sun is significant, 
we will carry out studies of rotating convection
in the future.
\par
Fig. \ref{back}
shows our reference state in comparison with Model S \citep{1996Sci...272.1286C}.
The reference stratification is determined by solving the
one-dimensional
hydrostatic equation and the realistic equation of state as:
\begin{eqnarray}
 && \frac{dp_0}{dr}=-\rho_0 g,\\
 && \rho_0 = \rho_0(p_0,s_0),\label{1dopal}\\
 && \frac{ds_0}{dr} = 0.
\end{eqnarray}
Eq. (\ref{1dopal}) is
calculated with the OPAL repository including partial ionization.
We set an adiabatic stratification as a reference and initial state. 
The
stratification becomes superadiabatic after 
convection develops as a consequence of radiative heating near the
bottom and radiative cooling at the top as described below.
The gravitational acceleration and the radiative diffusion are adopted
from Model S.
The boundary is set at $r=0.998R_\odot$ with the values
from Model S, and the equations are
integrated inward. 
\par
On the actual Sun, the surface is
continuously cooled by the radiation. 
Since our boundary is not located on  the real solar surface (even
though it is unprecedentedly closer to the real surface), we add
an artificial cooling ($\Gamma$) as:
\begin{eqnarray}
\Gamma(r) = -\frac{1}{r^2}\frac{\partial }{\partial r}(r^2 F_\mathrm{s}),
\end{eqnarray}
where
\begin{eqnarray}
&& r^2F_\mathrm{s}(r) =
 r^2_\mathrm{min}F_\mathrm{r}(r_\mathrm{min})
 \exp\left[-\left(\frac{r-r_\mathrm{max}}{d_\mathrm{c}}\right)^2\right],\label{cooling}\\
&&F_\mathrm{r}(r) = -\kappa_\mathrm{r}\rho_0 c_\mathrm{p}\frac{dT_0}{dr}\label{radiative},
\end{eqnarray}
where $r_\mathrm{min}$ and $r_\mathrm{max}$ denote the location of the
bottom and the top boundary, respectively.
This procedure ensures that the radiative luminosity imposed at the
bottom is released through the top boundary. Since the
realistic simulation for the near-surface layer shows that the thickness
of the cooling layer by the radiation is similar to the local pressure
scale height \citep[e.g. ][]{2009ASPC..416..421S}, we adopt two pressure
scale heights for the thickness of the cooling layer, i.e.,
$d_\mathrm{c}=2H_{p0}(r_\mathrm{max})$ except for cases H1 and M1 (see table
\ref{param}), where $H_{p0}=p_0/(\rho_0g)$ is the
pressure scale height. \par
We carry out three calculations named H0, H1 and H2, with different settings that are primarily
hydrodynamic. In case H0, the top boundary is located at
$r=0.99R_\odot$, and the density contrast 
$\rho_0(r_\mathrm{min})/\rho_0(r_\mathrm{max})$ exceeds 600.
To our knowledge, this is the largest value
achieved so far in numerical simulations of solar global convection.
In cases H1 and H2, the top
boundary is at $r=0.96R_\odot$, and the density contrast is around 40.
In case H1, the thickness of the cooling layer is the same as 
case H0, i.e., $d_\mathrm{c}=3740\ \mathrm{km}$, which is the two pressure
scale heights at $r=0.99R_\odot$, while case H2 adopts two scale
heights at its top boundary ($r_\mathrm{max}=0.96R_\odot$) for the
thickness ($d_\mathrm{c}$). 
Our magnetic runs M0, M1 and M2 use the same settings and differ only
through the initial magnetic field added.
We note that the thickness of the cooling layer
($d_\mathrm{c}$) has almost the same role as the value of the turbulent
thermal diffusivity on the entropy that is adopted in the ASH simulation
\citep{2000ApJ...532..593M,2008ApJ...673..557M}.
\par
In this study, the factor of the RSST is set to make the adiabatic
reduced speed of sound uniform in space. The adiabatic speed of sound is defined
as:
\begin{eqnarray}
 c_\mathrm{s}(r) = \sqrt{\left(\frac{\partial p}{\partial \rho}\right)_s}.
\end{eqnarray}
Then the factor of the RSST is set as:
\begin{eqnarray}
 \xi(r) = \xi_0 \frac{c_\mathrm{s}(r)}{c_\mathrm{s}(r_\mathrm{min})}.
\end{eqnarray}
In this study, we adopt $\xi_0=120$ for all the calculations, thus the
reduced speed of sound
${\hat{c}_\mathrm{s}} \equiv c_\mathrm{s}/\xi= 1.88\ \mathrm{km\ s^{-1}}$ at all depths.
\cite{2012A&A...539A..30H} suggest that the
RSST is valid for the thermal convection under the criterion of 
$v_\mathrm{rms}/\hat{c}_\mathrm{s} < 0.7$,
where $v_\mathrm{rms}$ is the RMS (root mean square) convection
velocity. Thus we can properly treat the convection with $v_\mathrm{rms}
< 1.3\ \mathrm{km\ s^{-1}}$ in this model. It is also suggested in
\cite{2012A&A...539A..30H} that the total mass is not conserved with the
inhomogeneous $\xi$, and we cannot avoid long-term drift from the
reference state, i.e., the mass continuously decreases or increases.
In this study, however, we adopt a different way to avoid
this type of long-term drift. When the equation of
continuity is treated as:
\begin{eqnarray}
 \frac{\partial}{\partial t}\left(\xi^2 \rho_1\right)
  = -\nabla\cdot\left(\rho_0 {\bf v}\right),
\end{eqnarray}
the value $\hat{M}$ is conserved in a rounding error with appropriate
boundary conditions, where 
\begin{eqnarray}
 \hat{M} = \int_V \xi^2 \rho_1 dV.
\end{eqnarray}
Although the radial distribution of the density is different from
the original one, the fluctuation part
remains small (e.g. $\rho_1/\rho_0 \sim 10^{-6}$), and this
does not affect the character of the thermal convection. It is
confirmed in \cite{2012A&A...539A..30H} that statistical features are
not influenced by the inhomogeneous $\xi$.
\par
We use the stress-free and the impenetrable boundary conditions for the
fluid velocity, $v_r$, $v_\theta$, and $v_\phi$. The free
boundary condition is adopted for the density and the entropy,
i.e., $\partial \rho_1/\partial r=\partial s_1/\partial r=0$. The
magnetic field is vertical at the top boundary, and the perfect conductor
boundary condition is used at the bottom boundary. 
\par
$\rho_1$, $B_r$, $B_\theta$, $B_\phi$ and $s_1$ are zero initially. The fluid
velocities $v_r$, $v_\theta$, and $v_\phi$ have small random
values. After the convection reaches a statistically steady state, a
uniform magnetic field ($B_\phi=100\ \mathrm{G}$) is added.
Although a net toroidal flux exists initially using this
condition, it disappears through the upper boundary
after around $75$ days,
since horizontal magnetic field is
zero there.
Thus we do not need to consider the influence of the net flux for our
local dynamo study.
\par
\begin{figure}[htbp]
 \centering
 \includegraphics[width=15cm]{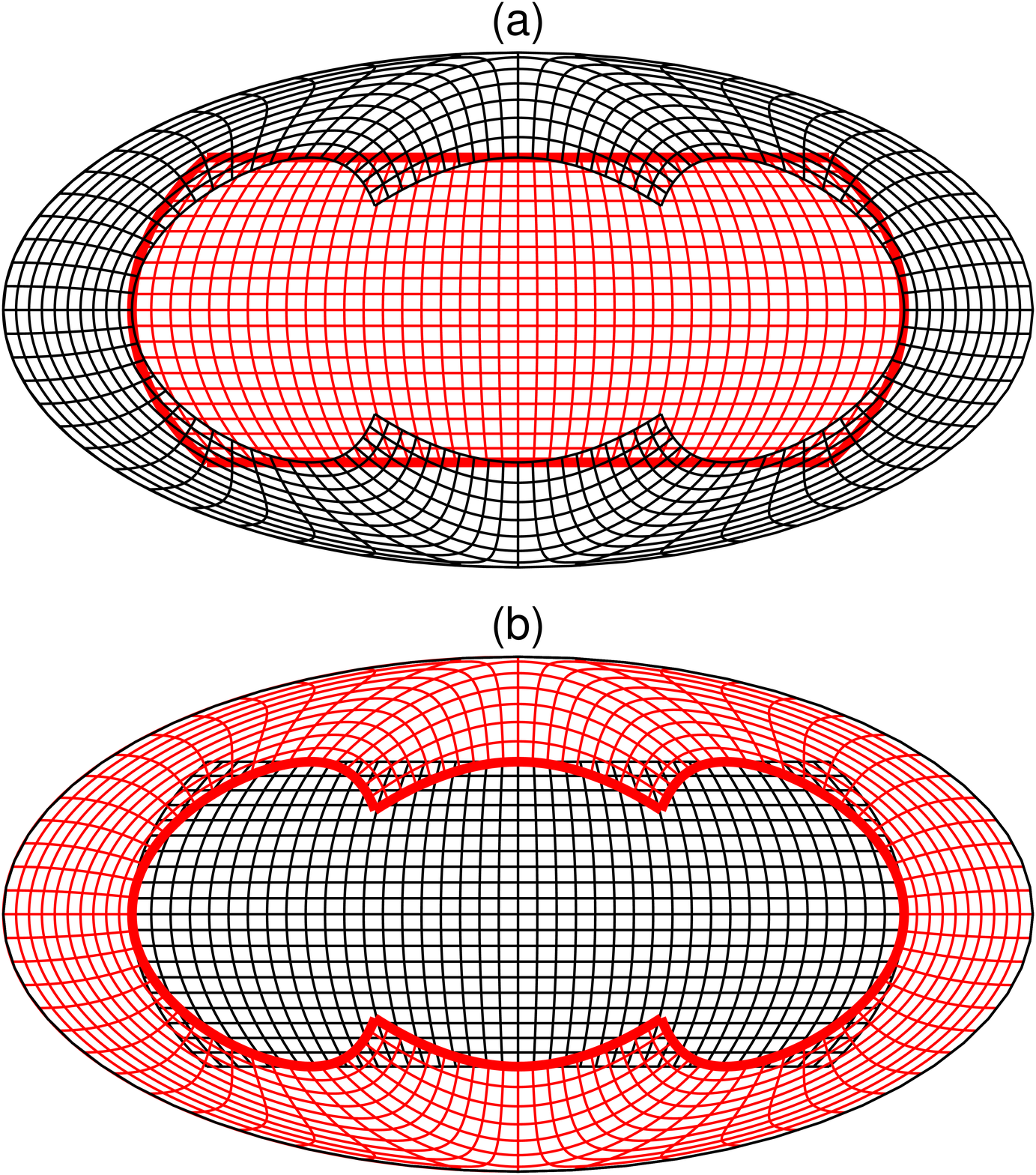}
 \caption{Red lines and black lines indicate the Yin and Yang grid,
 respectively. (a) and (b) show the geometry on the mollweide
 projection from the different viewpoints. The thick red lines show the
 boundaries for both the Yin and Yang grids.
\label{yinyang}}
\end{figure}

Under the setting explained above, we solve the equations
(\ref{eqco})-(\ref{eqen}) by using the fourth-order Runge-Kutta scheme for the
time-integration and the fourth-order space-centered derivative
\citep{2005A&A...429..335V}. 
The divergence free condition, i.e., $\nabla\cdot {\bf B}=0$, is
maintained with the diffusion scheme for each Runge-Kutta loop
(see the \cite{2009ApJ...691..640R} appendix for details).
In order to include all the spherical shell, we adopt the Yin-Yang grid
\citep{2004GGG.....5.9005K}. The Yin-Yang grid is a set of two congruent
spherical geometries. They are combined in a complemental way to cover a
whole spherical shell. The boundary condition for each grid is
calculated using the interpolation of the other grid. 
In all of our calculations, each grid covers
$0.715R_\odot<r<r_\mathrm{max}$,
$\pi/4-\delta_\theta<\theta<3\pi/4+\delta_\theta$, and 
$-3\pi/4-\delta_\phi <\phi<3\pi/4+\delta_\phi$, where $\delta_\theta$
and $\delta_\phi$ are the margins for
the interpolation. Here $\delta_\theta=3\Delta\theta/2$ and
$\delta_\phi=3\Delta\phi/2$ are adopted for the
interpolation with the third-order function, where $\Delta\theta$ and
$\Delta\phi$ are the grid spacing in the latitudinal and longitudinal
directions.
Here $r=0.715R_\odot=r_\mathrm{min}$ is
the location of the base of the convection zone and $r_\mathrm{max}$ is
taken as a parameter ($r_\mathrm{max}=0.99R_\odot$ for the typical case).
Although the boundary condition for the Yin grid is applied on the
edge of the Yin grid, the boundary condition for the Yang grid is
applied on the edge of the Yin grid in order to avoid the double
solution in the overlapping area of the Yin-Yang grid. Fig. \ref{yinyang}
shows this configuration. The thick red lines show the location of the
horizontal boundary for both Yin and Yang grids.
\par
The
horizontal grid spacing is 1,100 km at the top boundary and 375
km radially. With the Yin-Yang geometry, the number of grid points is 
$1024 (N_\theta)\times 3072 (N_\phi)\times 2$. The last factor 2
indicates a Yin and Yang pair. The number of grid points in the radial direction is
shown in Table \ref{param}.
Since this resolution in Yin-Yang geometry has almost the same quality
as that of $512(r)\times 2048(\theta)\times 4096(\phi)$ in ordinary spherical geometry
in cases H0 and M0, we
succeed in doubling the resolution in each direction from the previous
study \citep{2008ApJ...673..557M}.
After the uniform magnetic field is added, the
calculations are newly named M0, M1 and M2, which use the results of H0,
H1, and H2, respectively.
Using a hybrid MPI and automatic intra-node parallelization approach,
the code scales efficiently up to $10^5$ cores with almost linear weak
scaling, and achieves a 14\% performance at maximum on the RIKEN
K-computer in Japan. Since our code includes almost no global
communication among cores, this linear scaling is expected to hold
further with larger core counts.
\par
\section{Results}\label{result}
\subsection{Structure of convection and the magnetic field}\label{structure}
\begin{figure}[htbp]
 \centering
 \includegraphics[width=15cm]{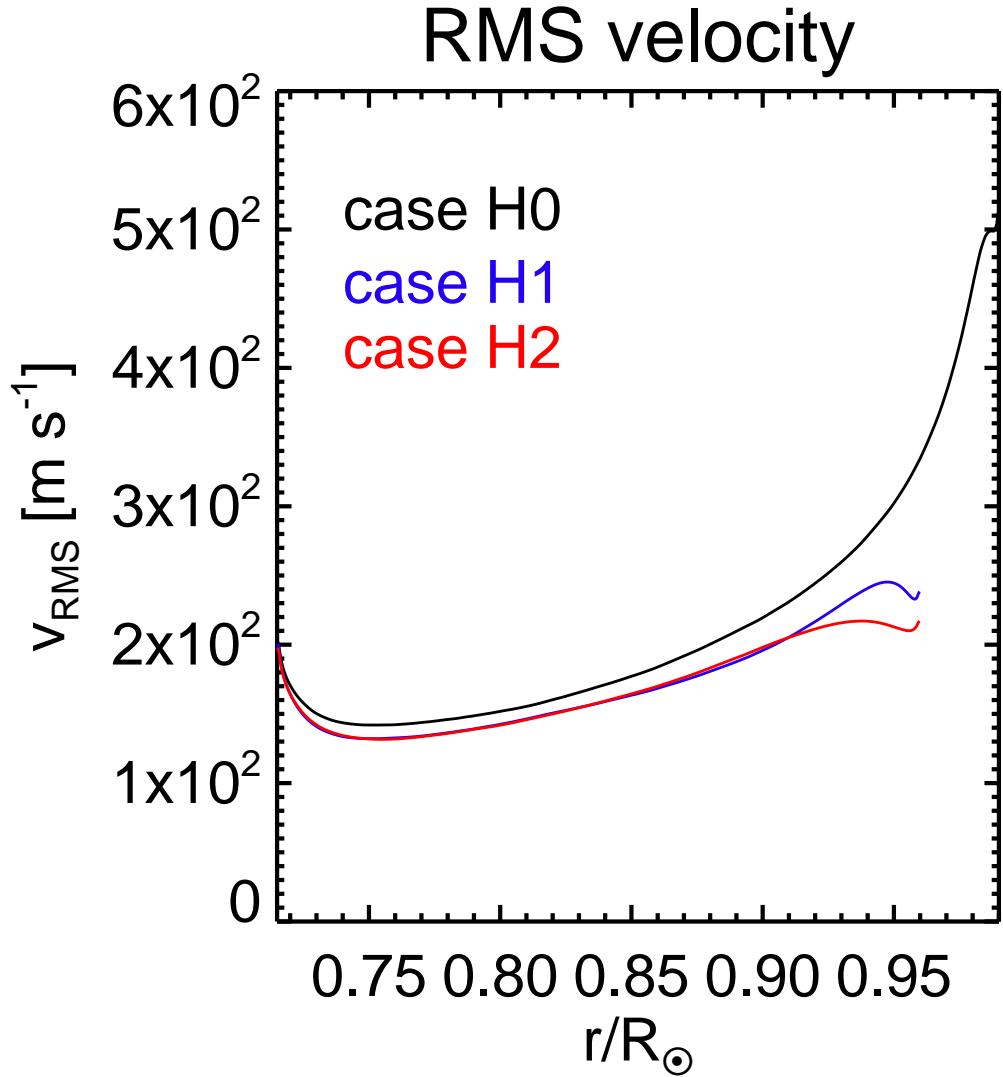}
 \caption{
 RMS (root mean square) velocities as a function of the depth are
 shown. The black, blue, and red lines show the results in cases H0,
 H1, and H2, respectively.
\label{RMS}}
\end{figure}

Fig. \ref{RMS} shows the RMS velocities in cases H0, H1, and H2. The
maximum RMS velocity is
$5\times 10^2\ \mathrm{m\ s^{-1}}$ at the top boundary in 
case H0.
Since the Mach number determined with the reduced speed
of sound $\hat{c}_\mathrm{s}$ is always under 0.3 throughout the
convection zone, the requirement for the
validity of the RSST in this study is well satisfied
\citep{2012A&A...539A..30H}.
The difference between cases H1 and H2 is mostly a local effect. The
thinner cooling layer in
case H1 leads to stronger acceleration of plasma and causes a higher RMS
velocity close to the top boundary.
In case H0 the RMS velocity is significantly larger in the near surface
layer ($>0.96R_\odot$), since
the energy has to be transported with a small background density. This
has some influence on the velocity
throughout the convection zone.
\par
\begin{figure}[htbp]
 \centering
 \includegraphics[width=15cm]{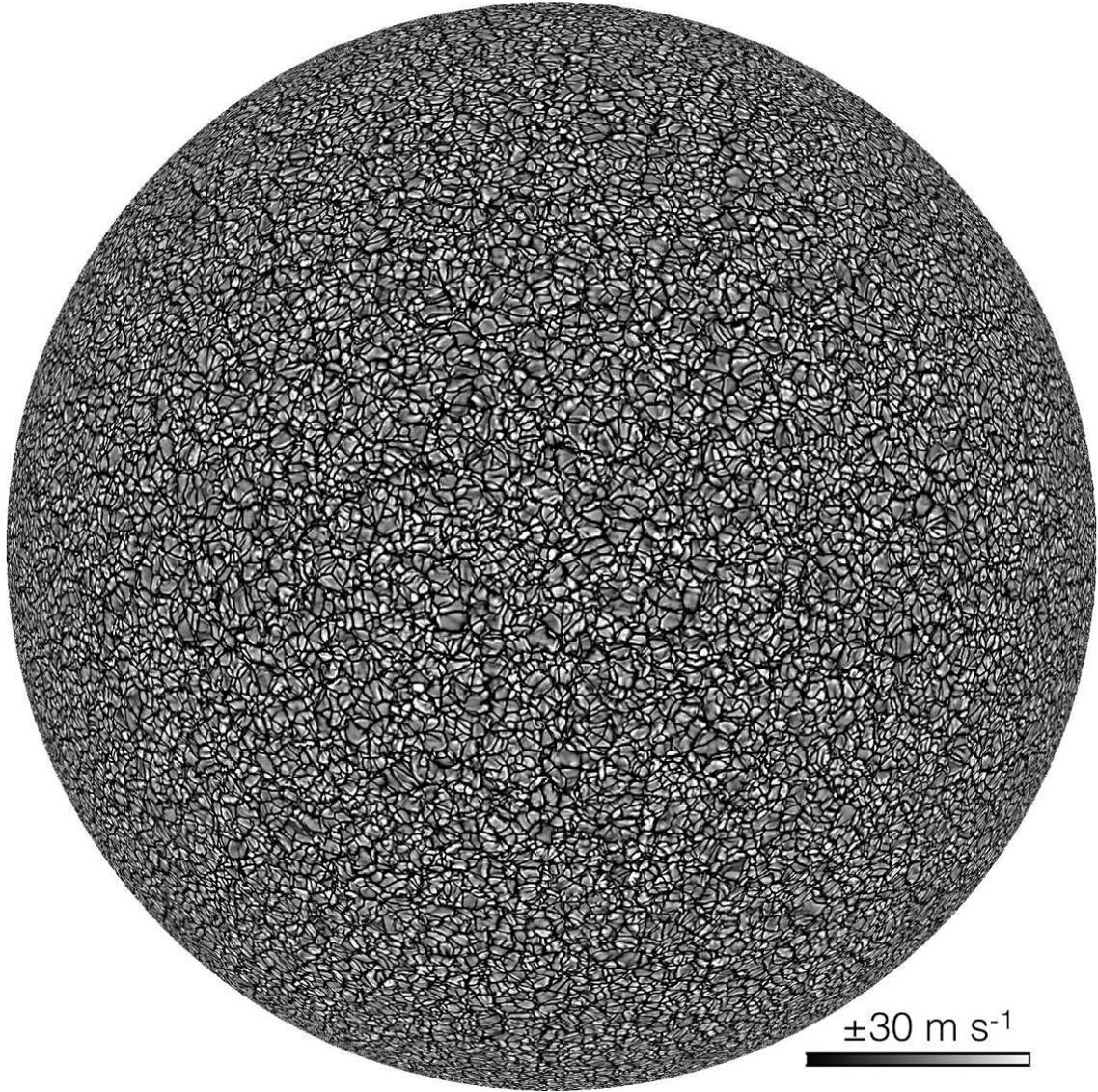}
 \caption{
The radial velocity ($v_r$) in case H0 on the orthographic
 projection. The movie is provided online.
\label{H0_pmap}}
\end{figure}
\begin{figure}[htbp]
 \centering
 \includegraphics[width=16cm]{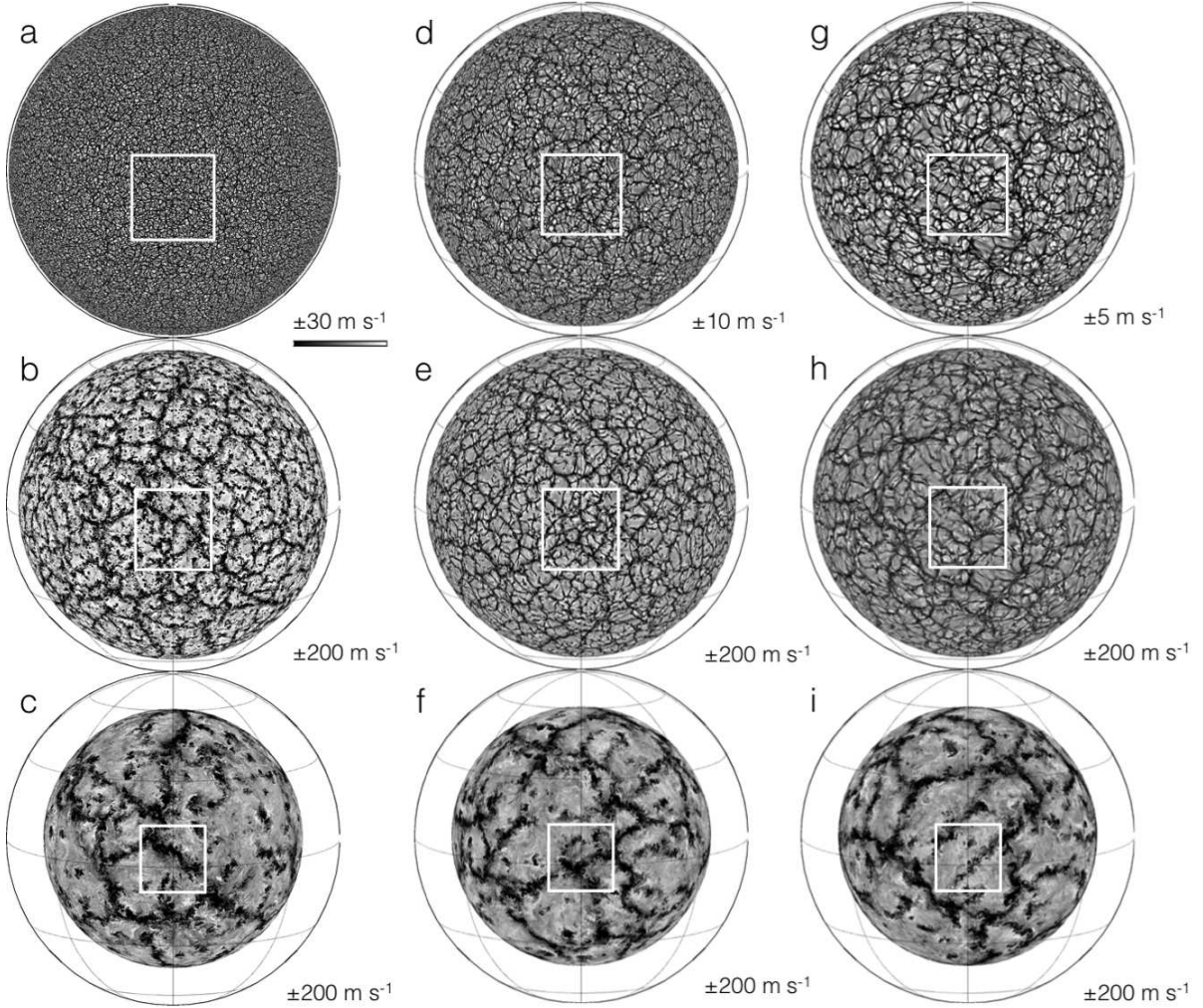}
 \caption{
 The radial velocities ($v_r$) are shown at $r=r_\mathrm{max}$ (a,d,g),
 $r=0.95R_\odot$ (b,e,h) and $r=0.85R_\odot$ (c,f,i).
The results in cases H0, H1, and H2 are shown in (a,b,c), (d,e,f),
 and (g,h,i), respectively. The black circle around each panel shows the
 location at $r=R_\odot$.
\label{H0H1_sp}}
\end{figure}

Fig. \ref{H0_pmap} shows the radial velocity ($v_r$) around the top
boundary in case H0
($r_\mathrm{max}=0.99R_\odot$, which is 7 Mm below the photosphere) in
the orthographic projection. 
(The corresponding movie is provided online.)
Note that since the shown location is close to the impenetrable top
boundary, the value of the radial velocity is rather small.
Since in the area near the upper boundary the pressure scale height is less than 2
Mm, the convection pattern shows significantly small cells of about
$\sim 7\ \mathrm{Mm}$. The typical cell size is slightly smaller
than the supergranulation that is observed in the photosphere. This study is the
first that well resolves the 10 Mm-scale convection pattern
in a calculation of the solar global convection zone.
Fig. \ref{H0H1_sp}a-c shows the radial
velocity at $r=0.99R_\odot$, $r=0.95R_\odot$, and $r=0.85R_\odot$ in 
case H0 by using the orthographic projection.
In deeper layers the
pressure scale height becomes larger (see Fig. \ref{back}e) and the
convection
cell increases in size.
A detailed analysis using
the spherical harmonic expansion of the convective structure is shown in
\S \ref{sh}. \par
\begin{figure}[htbp]
 \centering
 \includegraphics[width=16cm]{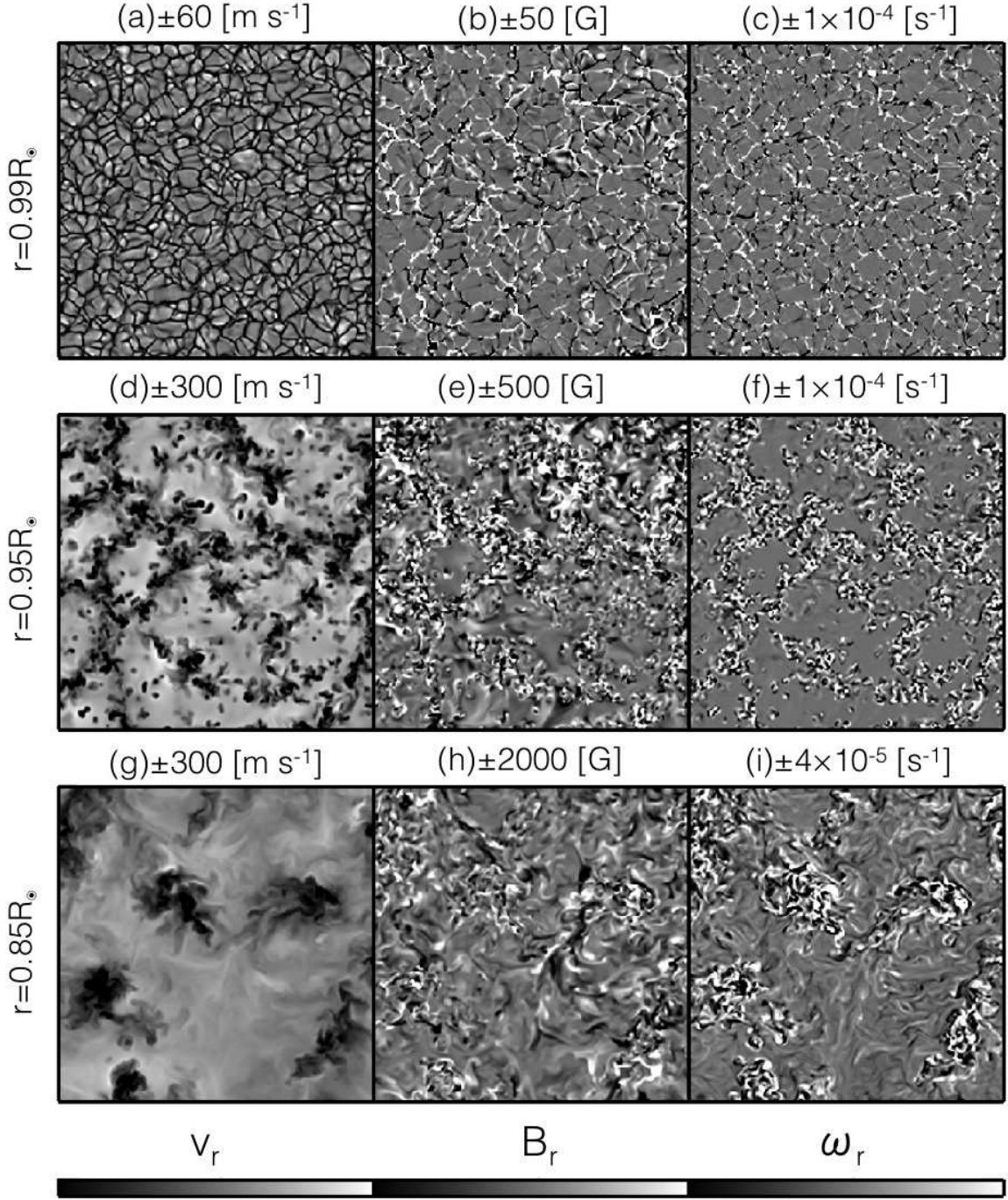}
 \caption{
The zoomed-in contour of the radial velocity ($v_r$: left panels), the
 radial magnetic field ($B_r$: middle panels) and the radial vorticities
 ($\omega_r$: right panels) at different depths. The field of view
 is $30^\circ$ both in the latitude and the longitude, which corresponds
 to the size of 370 Mm at the top boundary.
\label{H0_multi}}
\end{figure}
\begin{figure}[htbp]
 \centering
 \includegraphics[width=16cm]{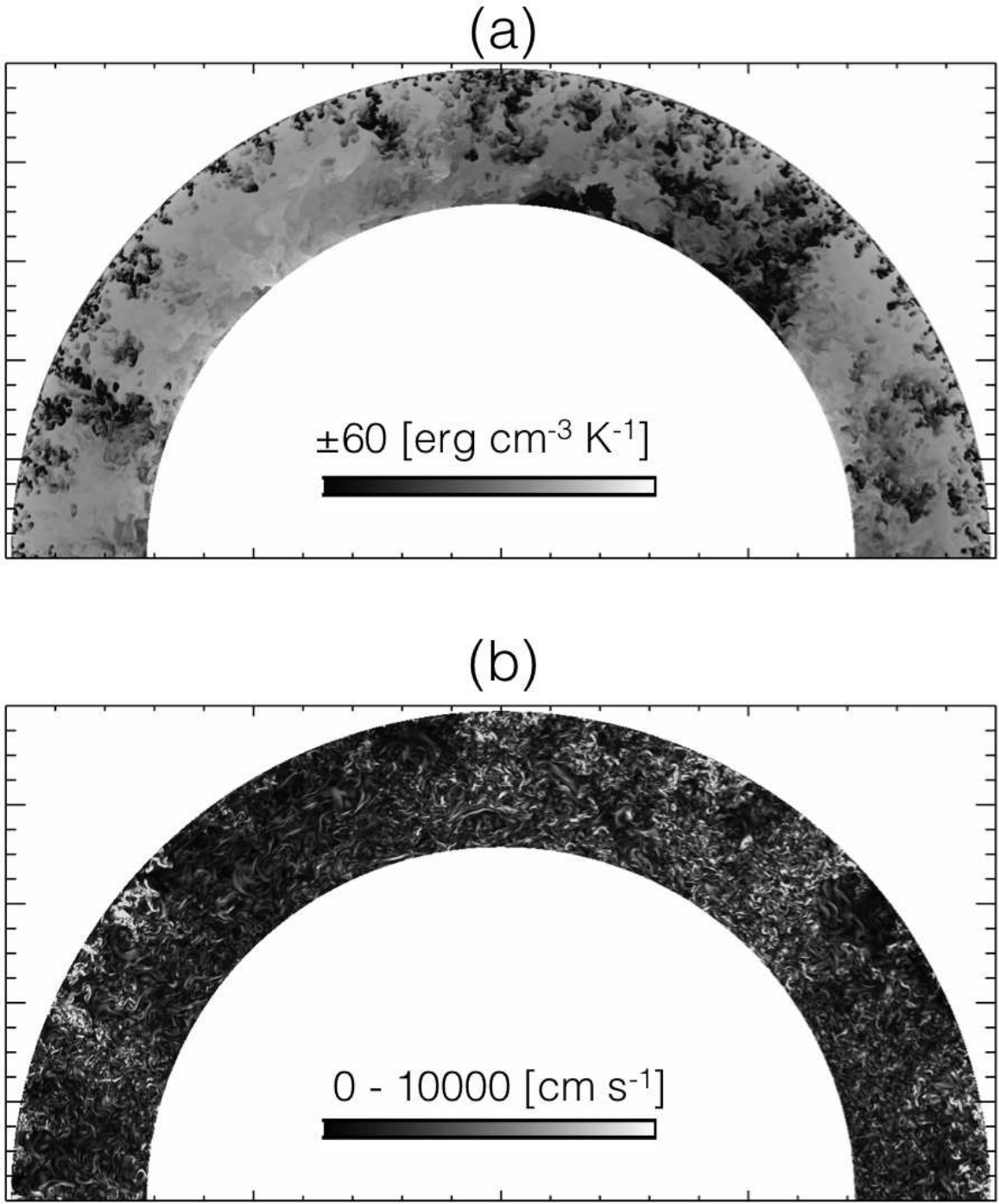}
 \caption{(a) $\rho_0[s_1-\langle s_1\rangle]$, and (b)
 $B/\sqrt{4\pi\rho_0}$ on the meridional plane at $\phi=0$,
 where $\langle{s}\rangle$ is the horizontal average of the entropy.
 \label{H0_meri}}
\end{figure}

Figs. \ref{H0_multi}a, d, and g show the zoomed-up contour of the radial
velocity in case M0, i.e., after the inclusion of the magnetic
field in H0. The region is indicated by the white rectangle in
Figs. \ref{H0H1_sp}a-c. Figs. \ref{H0_multi}c, f, and i show the
vorticity ($\omega_r=(\nabla\times {\bf v})_r$) at
$r=0.99R_\odot$, $0.95R_\odot$, and $0.85R_\odot$,
respectively. As already seen in case H0 (Fig. \ref{H0H1_sp}), it is clear that the scale
of the thermal convection pattern significantly depends on the depth. 
In addition, the large-scale downflow is associated with the small-scale and
strong vorticity in the deeper layer (especially in
$r=0.85R_\odot$). 
Fig. \ref{H0_meri}a shows
$\rho_0[s_1(r,\theta,\phi=0)-\langle s_1\rangle]$ in the meridional
plane, where $\langle s_1 \rangle$ is the horizontal average of the
entropy. The low
and high entropy materials correspond to the downflow and upflow,
respectively. In the
near-surface region, the convection structure shows 
a combination of broad upflows and
narrow downflows $\sim 7\ \mathrm{Mm}$ forming at the top boundary. These
small-scale downflows merge in the middle of the convection
zone and build large-scale downflow. Although the overall structure of
such convection is large, there is a superimposed turbulent pattern especially in
the downflow region, which is shown in Fig. \ref{H0_multi}.
\par
\begin{figure}[htbp]
 \centering
 \includegraphics[width=12cm]{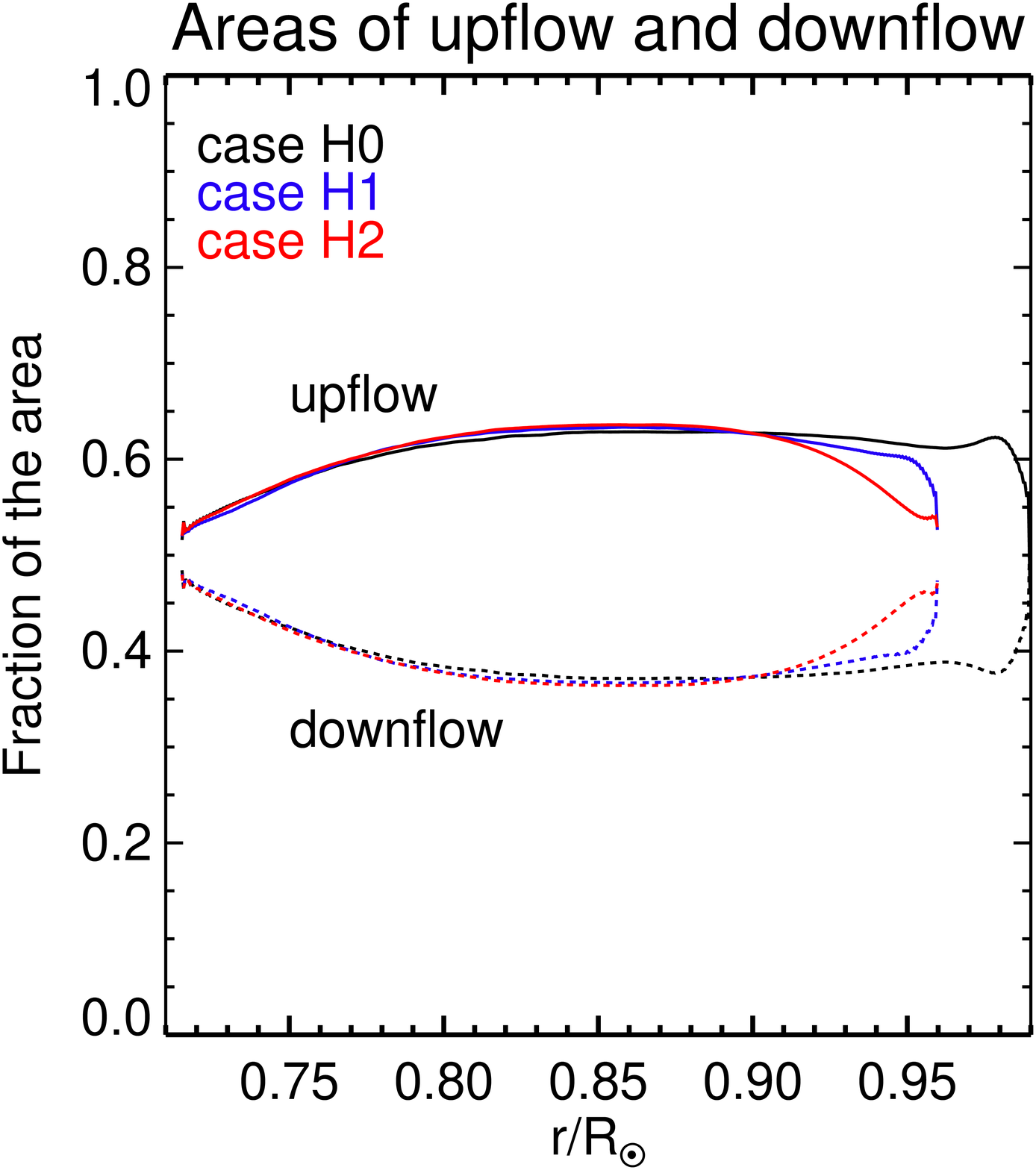}
 \caption{The occupied fraction of the area of the upflow (solid line) and downflow
 (dashed line). The black, blue and red lines show the results in 
 cases H0, H1, and H2, respectively.
 \label{area}}
\end{figure}

Figs. \ref{H0H1_sp}d, e, and f show the results of case H1 with
a different location of the top boundary
($r_\mathrm{max}=0.96R_\odot$) at $r=r_\mathrm{max}$, $r=0.95R_\odot$,
and $r=0.85R_\odot$, respectively. Since the location and the pressure
scale height of the shown images is different between case H0 in Figs. \ref{H0H1_sp}a
($r=r_\mathrm{max}=0.99R_\odot$) and H1 in d ($r=r_\mathrm{max}=0.96R_\odot$),
it is natural that the convective structures are much
different, i.e., those in H0 have smaller scale convection than in H1.
It is more important that the structures at the same depth, $0.95R_\odot$,
are significantly different from each other between these cases (H0 in
Fig. \ref{H0H1_sp}b and H1 in Fig. \ref{H0H1_sp}e. 
The small-scale
downflow plumes penetrate near the surface layer
and influence its
structure (see also Fig. \ref{H0_meri}a). When the downflow goes deeper,
the influence becomes
smaller. This is seen in the comparison of Figs. \ref{H0H1_sp}c 
($r_\mathrm{max}=0.99R_\odot$ and $r=0.85R_\odot$) and f
($r_\mathrm{max}=0.96R_\odot$ and $r=0.85R_\odot$), where
the difference of convection structure seems insignificant.
Figs. \ref{H0H1_sp}g, h, and i show the results in case H2, in which
the location of the top boundary is the same as in H1 but
the thickness of the cooling layer is larger.
The convective structure around the top boundary shows the largest scale
(Fig. \ref{H0H1_sp}g), while
again the difference becomes smaller in the deeper layer (Fig. \ref{H0H1_sp}c,
f, and i).
This is also shown in Fig. \ref{area} by 
the areas occupied by the upflow and the
downflow, i.e., positive and negative radial velocities ($v_r$).
Up to the middle of the convection zone ($<0.9R_\odot$), 
cases H0, H1 and H2 all show similar behavior, i.e., 
the fractional area by the upflow is larger ($\sim 65\%$) but decreases
below $\sim 0.85 R_\odot$ to equal to that of the downflow.
Interestingly, this behavior is quantitatively the same in spite of the significant
difference between H0 and H1 in the density contrast (see table
\ref{param}).\par
\subsection{Integrated energy flux }\label{integrate_flux}
\begin{figure}[htbp]
 \centering
 \includegraphics[width=12cm]{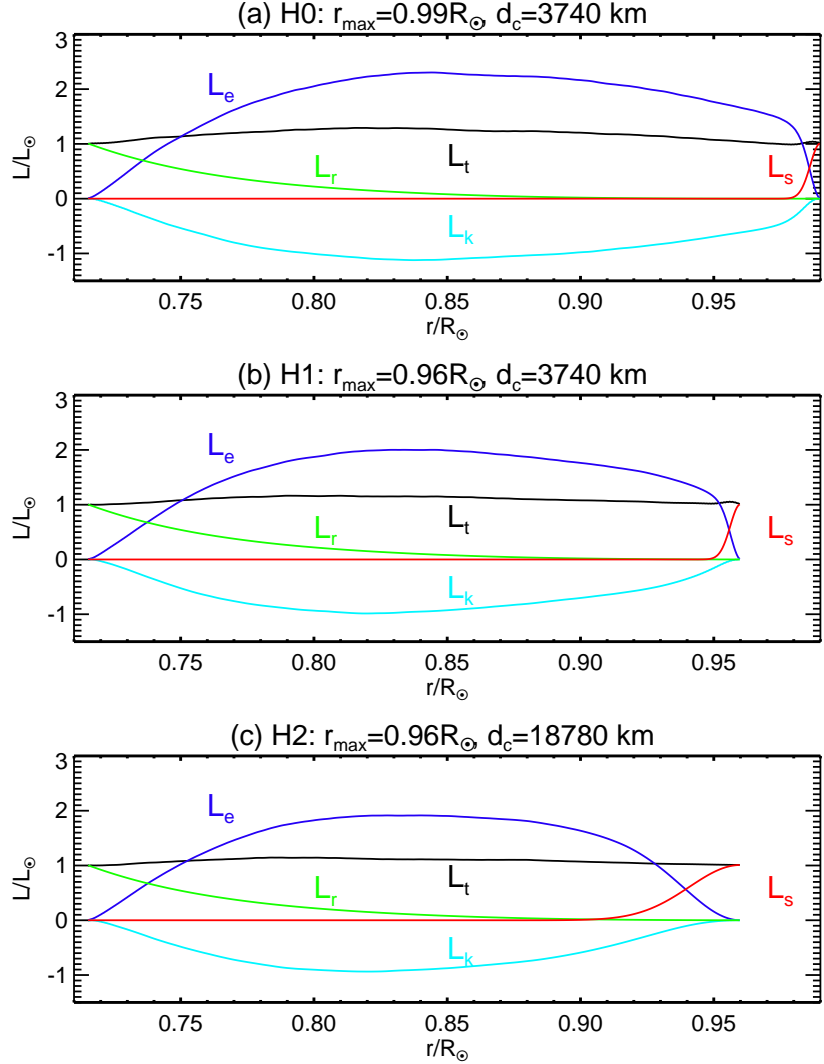}
 \caption{The integrated fluxes are shown. Panels a, b, and c show
 the results in cases H0, H1, and H2, respectively. The black, blue, green, red,
 and light blue lines show the total, enthalpy, radiative, surface cooling,
 and kinetic fluxes, respectively.
 \label{flux}}
\end{figure}

Fig. \ref{flux} shows the integrated fluxes. 
The integrated enthalpy flux ($L_\mathrm{e}$), the integrated kinetic
flux ($L_\mathrm{k}$), the radiative luminosity ($L_\mathrm{r}$) and the
luminosity form of the surface cooling ($L_\mathrm{s}$) are defined as:
\begin{eqnarray}
 L_\mathrm{e} &=& \int_s
  \left[\rho_0e_1 + p_1 -
   \frac{p_0\rho_1}{\rho_0}\right]v_r dS,\\
 L_\mathrm{k} &=& \int_s
 \frac{1}{2}\rho_0v^2 v_r dS,\\
L_\mathrm{r} &=& \int_s F_\mathrm{r}(r) dS,\\
L_\mathrm{s} &=& \int_s F_\mathrm{s}(r) dS.
\end{eqnarray}
The radiative flux ($F_\mathrm{r}$) and surface cooling flux
($F_\mathrm{s}$) are defined in eq. (\ref{cooling}) and
eq. (\ref{radiative}), respectively.
The derivation of the enthalpy flux is given
in Appendix \ref{enthalpy_flux}. Fig. \ref{flux}a shows the integrated
fluxes in case H0 ($r_\mathrm{max}=0.99R_\odot$ and
$d_\mathrm{c}=3740\ \mathrm{km}$). The total integrated flux 
$L_\mathrm{t}=L_\mathrm{e}+L_\mathrm{k}+L_\mathrm{r}+L_\mathrm{s}$ is
almost constant along the depth. This indicates that the convection zone in
our calculation is in energy equilibrium. This takes
about 50 days.
Note that since we do not use a conservative form for
the total energy nor estimate the energy flux contributions caused by
the artificial diffusivities, the total flux is not completely constant.
The enthalpy flux
transports twice the solar luminosity upward at maximum, and the kinetic
flux transports almost the same amount of the energy as the solar
luminosity downward at maximum. Although the kinetic energy flux is
frequently ignored in
the one-dimensional mixing length model
\citep[e.g.][]{2004suin.book.....S}, our result
shows the importance of the
kinetic flux. This has been already suggested by
\cite{2008ApJ...673..557M}. 
Figs. \ref{flux}b, and c show the results in cases H1 and H2,
respectively. 
The integrated fluxes show almost the similar behavior as those in 
case H0, but the maximum absolute values of the enthalpy and kinetic
flux are
smaller. Since these absolute values gradually decrease from
H0 to H2, we conclude that both the thickness of the cooling layer and
the location of the upper boundary contribute to this result.
It is possible that when the upper
boundary becomes closer to the real solar surface and the cooling layer
becomes thinner, the absolute values of the enthalpy flux and the kinetic
flux become even larger than our case H0.\par
\begin{figure}[htbp]
 \centering
 \includegraphics[width=12cm]{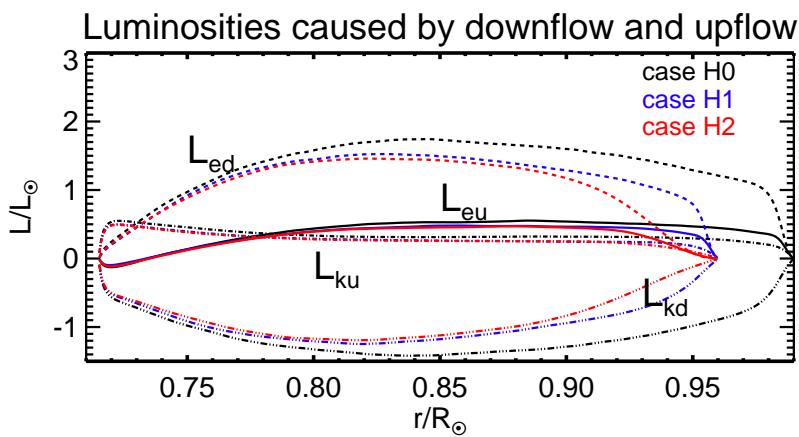}
 \caption{ The enthalpy and kinetic flux transported by the upflow
 ($L_\mathrm{eu}$ and $L_\mathrm{ku}$) and the downflow ($L_\mathrm{ed}$
 and $L_\mathrm{eu}$) are shown. The black, blue and red lines show the
 results in cases H0, H1, and H2, respectively. The solid and dashed
 lines indicate the enthalpy flux in the upflow and the downflow, respectively,
 and the dash-dot and dash-dot-dot-dot lines indicate the kinetic energy
 flux in the upflow and the downflow, respectively.
 \label{aflux}}
\end{figure}

Fig. \ref{aflux} shows the integrated enthalpy flux and the kinetic flux transported by
upflow ($L_\mathrm{eu}$, $L_\mathrm{ku}$) and downflow ($L_\mathrm{ed}$,
$L_\mathrm{kd}$), respectively. 
Note that the enthalpy flux by upflow and downflow is
estimated with the perturbation from the reference state.
Regarding the enthalpy flux,
both upflow and downflow transport energy upward. The downflow
transports most of the energy ($>70\%$). We note that the enthalpy flux of
the upflow shows the negative value near the bottom boundary, since the cool
fluid bounces at the boundary and moves upward. Regarding the kinetic energy flux,
upflows (downflows) transport energy upward (downward). The larger kinetic energy flux of the downflows make the
kinetic energy flux negative. These results show that
downflows play a key role in the transports of energy in the solar convection zone.
While we find a significant
difference in the overall amplitude among cases H0 to H2, the ratio of contributions from up- and
downflows does not change much despite the significant difference in the density contrast.
\subsection{Analysis using spherical harmonics for the hydrodynamic cases}\label{sh}
In this section, the results of the analysis using the spherical
harmonics expansion are shown. We focus
on the question: How do the
location of the upper boundary and the thickness of the surface cooling
layer influence the convective structure throughout the convection zone?\par
A real
function $f(\theta,\phi)$ can be expressed in spherical harmonics as
\begin{eqnarray}
 f(\theta,\phi) = \sum^{l_\mathrm{max}}_{l=0}\sum^l_{m=-l} f_{lm}Y_{lm}(\theta,\phi),
\end{eqnarray}
where $Y_{lm}(\theta,\phi)$ is the spherical harmonics for degree $l$
and order $m$. In the analysis
the absolute total value along $m$ without $m=0$
\begin{eqnarray}
 \bar{f}_l = \sqrt{\sum_{m=-l}^l |f_{lm}|^2},
\end{eqnarray}
is shown. The value is normalized in order to satisfy the relation:
\begin{eqnarray}
 \frac{\int_\Omega (f(\theta,\phi)-\langle
  f(\theta,\phi)\rangle)^2\sin\theta d\theta d\phi}{4\pi}
   = \sum_{l=1}^{l_\mathrm{max}} \bar{f}_l^2.
\end{eqnarray}
Our spherical harmonic analysis is performed using the
freely available software archive SHTOOLS (shtools.ipgp.fr).
\par
\begin{figure}[htbp]
 \centering
 \includegraphics[width=16cm]{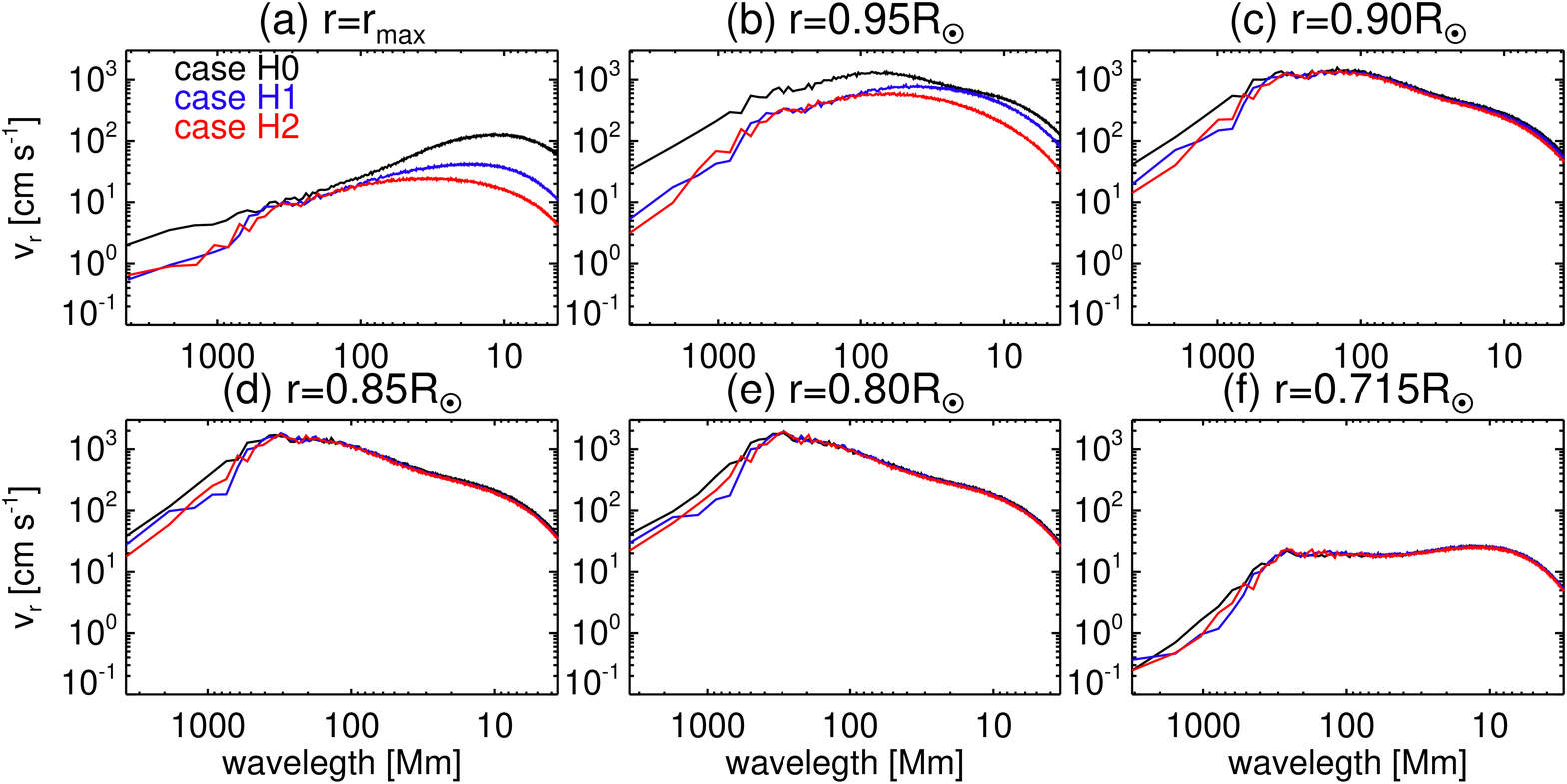}
 \caption{ Spectra of the radial velocity at
 (a) $r=r_\mathrm{max}$,
 (b) $r=0.95R_\odot$, (c) $r=0.90R_\odot$, (d) $r=0.85R_\odot$, 
 (e) $r=0.80R_\odot$, (f) $r=0.715R_\odot$.
 The black, blue, and red lines specify the results
 in cases H0, H1, and H2, respectively.
 \label{vr}}
\end{figure}
\begin{figure}[htbp]
 \centering
 \includegraphics[width=16cm]{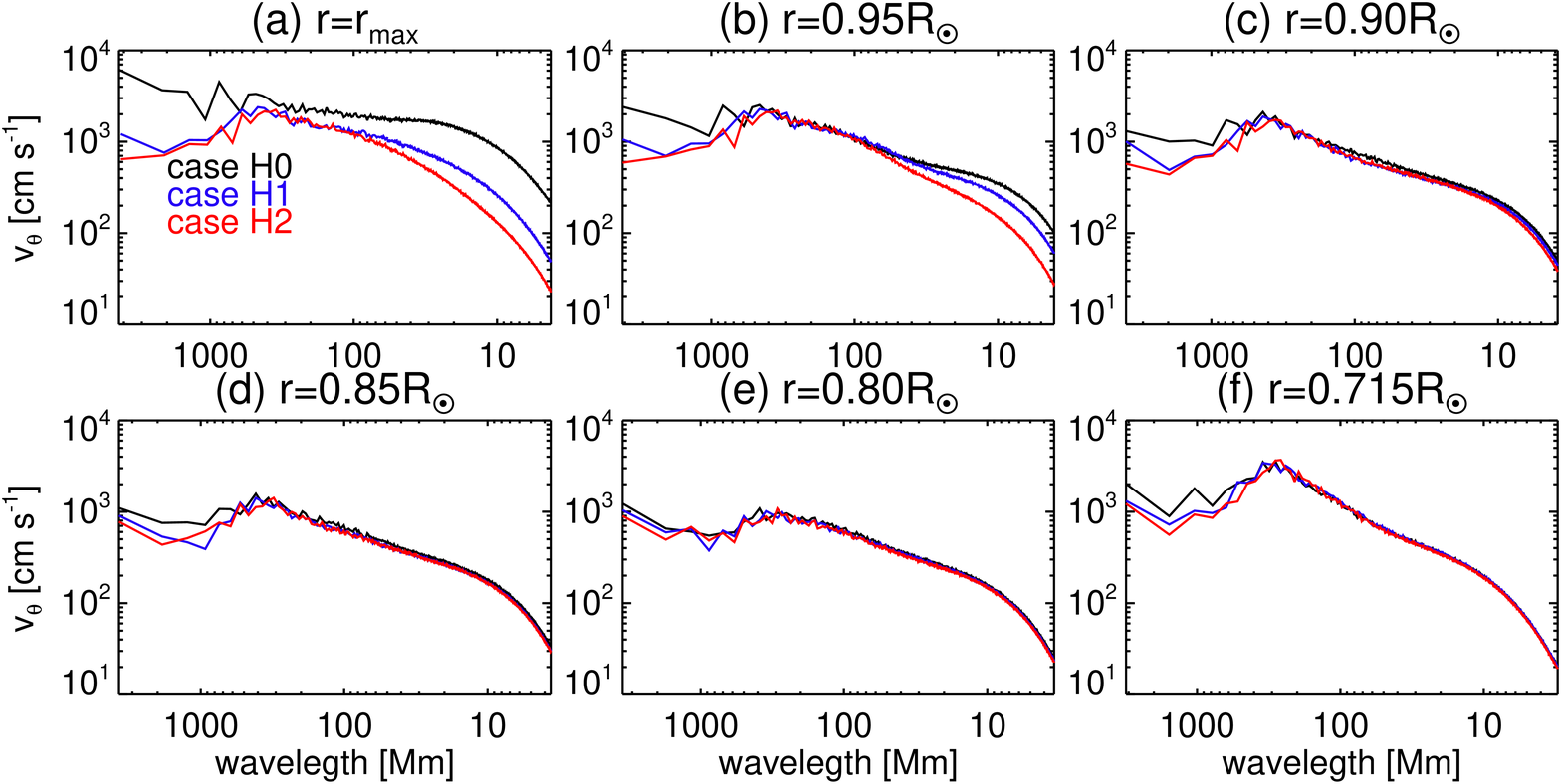}
 \caption{ Spectra of the latitudinal velocity at
 (a) $r=r_\mathrm{max}$,
 (b) $r=0.95R_\odot$, (c) $r=0.90R_\odot$, (d) $r=0.85R_\odot$, 
 (e) $r=0.80R_\odot$, (f) $r=0.715R_\odot$.
 The black, blue, and red lines specify the results
 in cases H0, H1, and H2, respectively.
 \label{vtheta}}
\end{figure}

Figs. \ref{vr} and \ref{vtheta} show the spectra of the
radial velocity
($v_r$) and the latitudinal velocity ($v_\theta$), respectively, as a
function of the horizontal wavelength ($L_\mathrm{h}=2\pi r/l$),
where $l$ is the spherical harmonic degree, i.e., the horizontal
wavenumber. 
The black, blue,
and red lines show the results in case 
H0 ($r_\mathrm{max}=0.99R_\odot$ and $d_\mathrm{c}=3740\ \mathrm{km}$),
H1 ($r_\mathrm{max}=0.96R_\odot$ and $d_\mathrm{c}=3740\ \mathrm{km}$), and
H2 ($r_\mathrm{max}=0.96R_\odot$ and $d_\mathrm{c}=18780\ \mathrm{km}$),
respectively. 
The black line in Fig. \ref{vr}a shows a peak around
$L_\mathrm{h}\sim7$-$8\ \mathrm{Mm}$ (see also Fig. \ref{H0_pmap}).
This peak
moves to the larger scale $L_\mathrm{h}$
with increasing depth. At $r=0.80R_\odot$, the peak is around
$L_\mathrm{h}\sim 300\ \mathrm{Mm}$ (Fig. \ref{vr}f, black line). 
This reflects the variation of the pressure scale height in the solar
model, $H_p = 1.9\ \mathrm{Mm}$ at $r=0.99R_\odot$ and 
$H_p = 44\ \mathrm{Mm}$ at $r=0.8R_\odot$ (see Fig. \ref{back}e).
Again the peak is in the smaller scale ($\sim 7\ \mathrm{Mm}$) at the
bottom boundary $r=r_\mathrm{min}=0.715R_\odot$. This is caused by the
collision between the downflow and the impenetrable bottom boundary. In
this process a thin boundary layer whose thickness is determined by
the numerical resolution is formed.
The scale of turbulence near the boundary
(Fig. \ref{vr}f) is a consequence of this boundary layer.
A comparison
of black lines in Fig. \ref{vr}a and \ref{vtheta}a indicates that the
peak of the latitudinal velocity is at a much larger scale ($\sim400\ \mathrm{Mm}$)
than the radial velocity, even close to the upper boundary.
We note that almost all the spectra have this feature
around $L_\mathrm{h}\sim 400\ \mathrm{Mm}$, i.e., the peak or the bended
feature in which the power-law index varies. 
The length of the scale $400\ \mathrm{Mm}$ is twice the
thickness of the convection zone, i.e., the radial extent of our
computational domain. This result is consistent with our previous study
\citep{2012ApJ...751L...9H}, which argues that the typical scale of the
convection is determined by the pressure (or density) scale height or the
height of the computational domain. Fig. \ref{vtheta}a shows that the
horizontal velocity is more likely to be influenced by the large scale structure.
\par
The influences from the location of the boundary
and the thickness of the cooling layer are discussed
by comparing the black, blue and red lines in Figs. \ref{vr}
and \ref{vtheta}. The common feature
is that the spectra do not depend on these two factors in the deep layer
($r\le 0.85R_\odot$:panels d, e, and f). This suggests that the
small-scale convection caused by the short pressure scale height or the
thin cooling layer cannot influence
the convection in the deeper region. 
On the top boundary (Figs. \ref{vr}a
and \ref{vtheta}a), the small-scale convection is suppressed gradually
from case H0 to H2. This is confirmation of our understanding
from the appearance of the convection pattern in \S \ref{structure}. In
the near-surface layer ($r=0.95R_\odot$), the difference still
remains unchanged. 
Moving the location of the top boundary upward increases the amplitude of the fluid velocity at
all scales (black line). Reducing the thickness of the cooling layer while keeping the location of the
boundary unchanged leads to an increase of  velocity on small scales (<50 Mm), while larger scales
are unaffected (blue line).
We point out that when the surface layer
($0.96R_\odot<r<0.99R_\odot$) is included, the spectrum of the
latitudinal velocity ($v_\theta$) is flat from the middle to the small
scale ($10<L_\mathrm{h}<40$ Mm in Fig. \ref{vtheta}b).
This is caused by the penetration of the corresponding-scale plume from
the near-surface layer. 

\subsection{Analysis using spherical harmonics and the probability
 density function for the magnetohydrodynamic cases}\label{sh_mag}
In this subsection, we analyze the results of the magnetohydrodynamic
calculation in cases M0, M1, and M2 using spherical harmonics
and probability density function. 
We note that cases M0, M1, and M2 use the same
parameters as cases H0, H1, and H2, respectively,  with the magnetic field.
We focus on the
influence of the location of the boundary layer and the thickness of the
cooling layer on the structure of the magnetic field. 
\par
\begin{figure}[htbp]
 \centering
 \includegraphics[width=16cm]{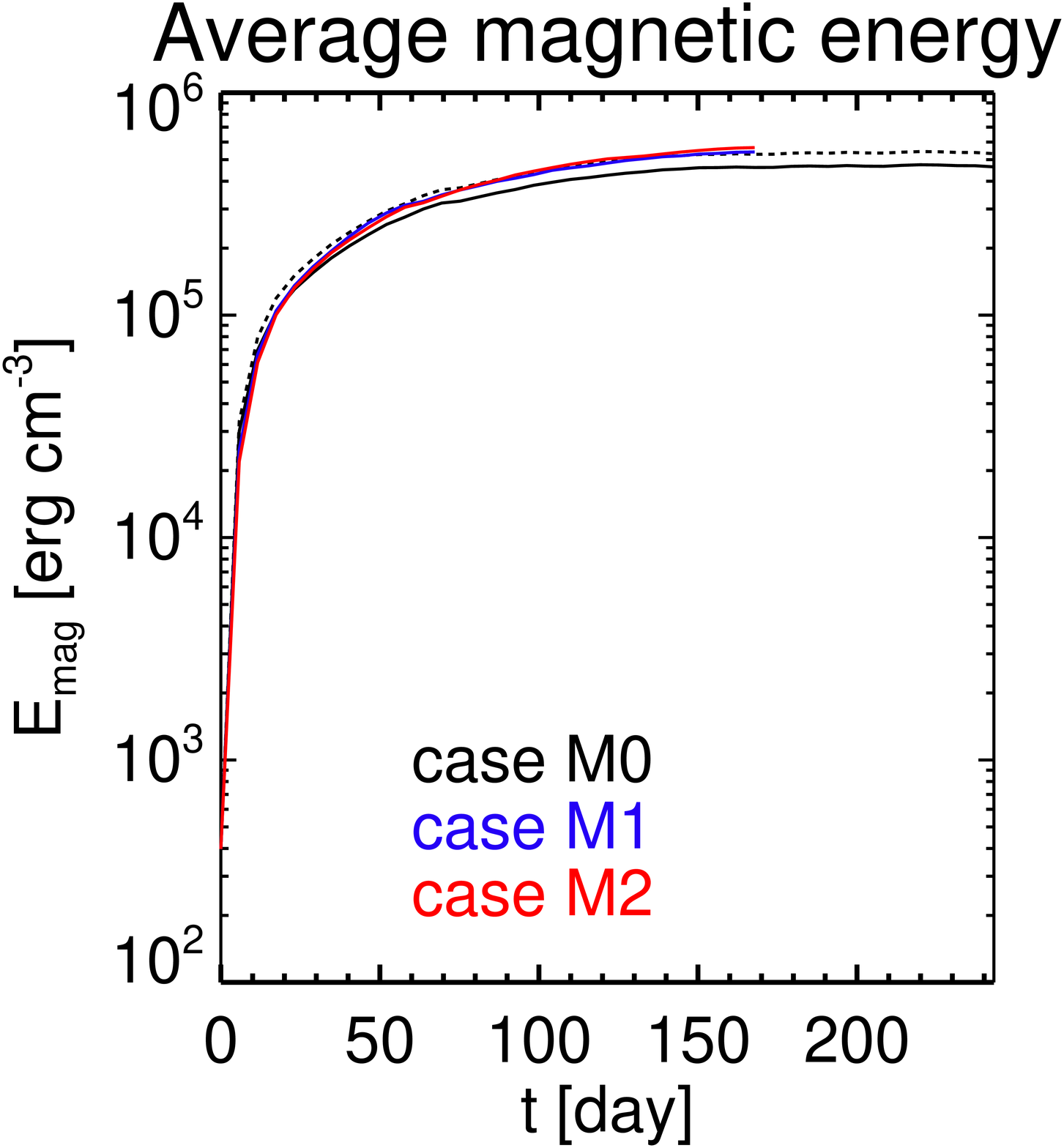}
 \caption{ The time evolution of the average magnetic energy is
 shown. The black, blue and red lines show the results in cases M0, M1
 and M2, respectively. The dashed line shows the magnetic energy in 
 case M0 averaged over $r_\mathrm{min}<r<0.96R_\odot$. $t=0$ is time
 at which the magnetic field is imposed.
 \label{magnetic_energy}}
\end{figure}

Fig. \ref{magnetic_energy} shows the time evolution of the
magnetic energy ($B^2/(8\pi)$) averaged over the simulation domain at
each time step. The initial linear growth stops
around 10 days after the input of the seed field in every case. Even
after that the magnetic energy
continues to increase gradually with a rather small growth
rate until it saturates after around 150 days in all
cases. A comparison among the cases is performed using the data from 
$t=115\ \mathrm{day}$ to $t=162\ \mathrm{day}$ in which the
generation of the magnetic field is almost saturated.
Basically, the differences among the cases are insignificant,
although case M0
saturates with a slightly smaller average magnetic energy. Since
the equipartition magnetic field strength in the near-surface region is smaller
due to the small density ($\rho_0$), the
increase of the volume in case  M0 causes the slight decrease in
the average magnetic energy. 
This conclusion can be supported by the fact that the average value over 
$r_\mathrm{min}<r<0.96R_\odot$ in M0 is larger, as shown by the dashed line in
Fig. \ref{magnetic_energy}.
The growth rate of this energy (dashed line) is higher
than those in M1 and M2 at $t<50$ days.
The time scale of the convection in the near-surface layer is
short. The generated
magnetic field is transported downward (see also the discussion about
the pumping in \S \ref{local_dynamo}). \par
\begin{figure}[htbp]
 \centering
 \includegraphics[width=16cm]{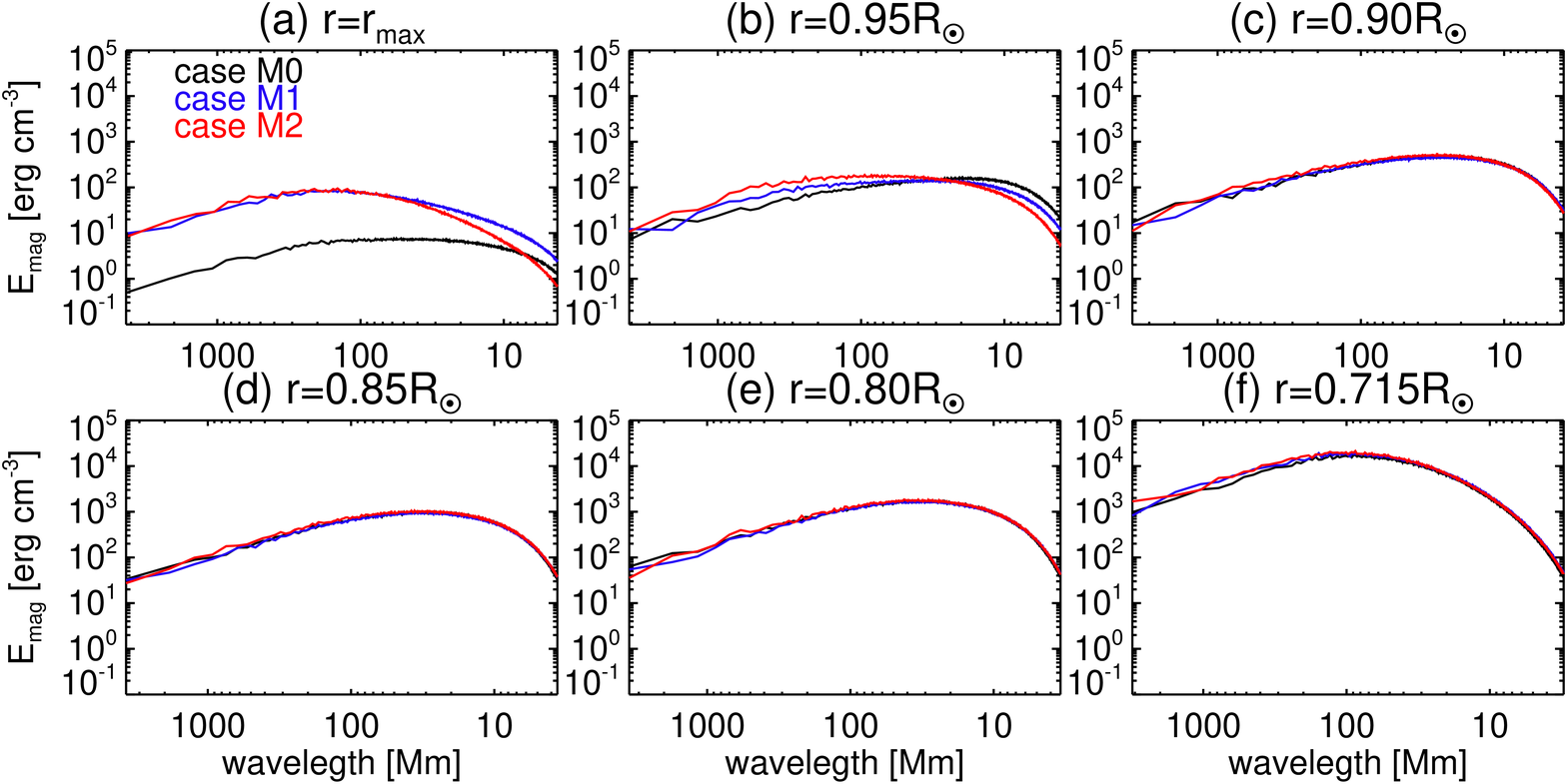}
 \caption{ Spectra of the magnetic energy ($B^2/8\pi$) at
 (a) $r=r_\mathrm{max}$,
 (b) $r=0.95R_\odot$, (c) $r=0.90R_\odot$, (d) $r=0.85R_\odot$, 
 (e) $r=0.80R_\odot$, (f) $r=0.715R_\odot$.
 The black, blue, and red lines specify the results
 in cases M0, M1, and M2, respectively.
 \label{magene}}
\end{figure}

Fig. \ref{magene} shows the spectra of the magnetic energy $B^2/(8\pi)$
at selected depths. Similar to the results introduced in the previous
sections, the differences can be seen in the near-surface layer, and these
differences become insignificant as we go to the deeper layers.
\par
\begin{figure}[htbp]
 \centering
 \includegraphics[width=16cm]{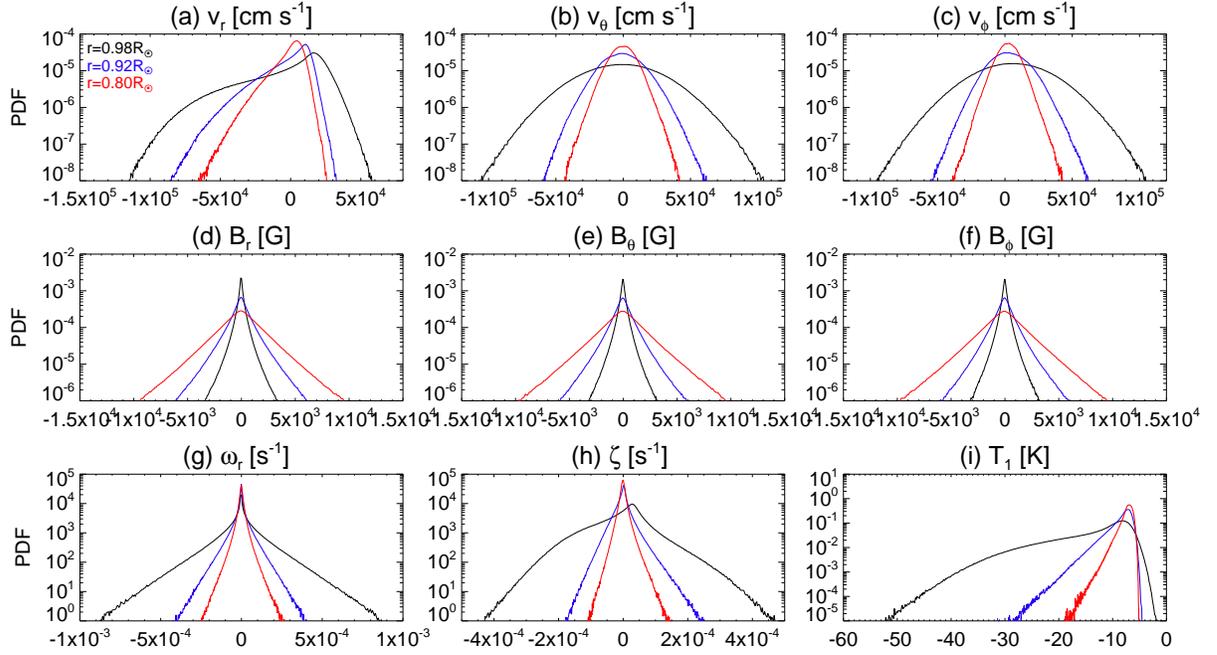}
 \caption{ The probability density function of (a) the radial
 velocity, (b) the latitudinal velocity, (c) the longitudinal velocity, (d)
 the radial magnetic field, (e) the latitudinal magnetic field, (f) the
 longitudinal magnetic field, (g) the radial vorticity, (h) the horizontal
 divergence, and (i) the temperature perturbation are shown using the
 results in case M0 at $t=115\ \mathrm{day}$.
 \label{pdf}}
\end{figure}
\begin{figure}[htbp]
 \centering
 \includegraphics[width=16cm]{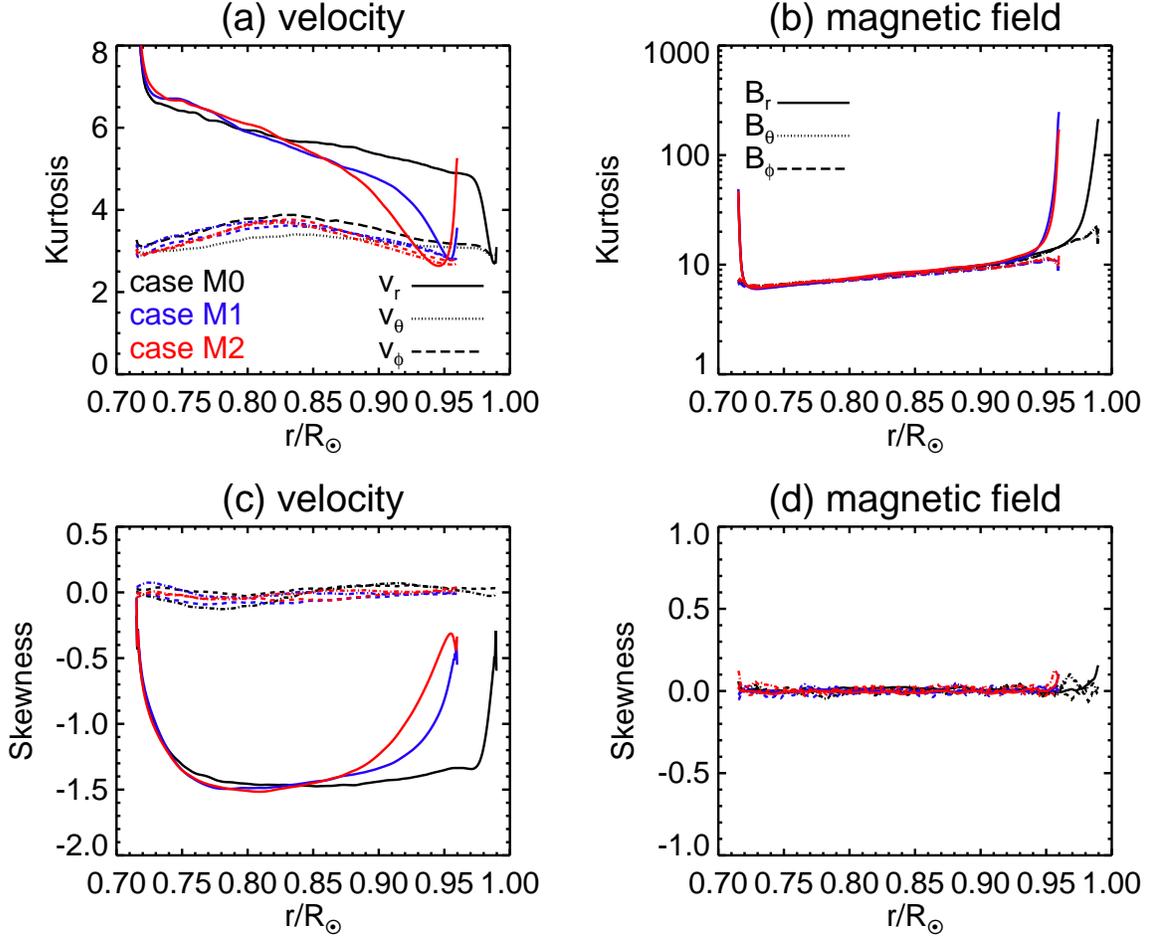}
 \caption{ The kurtosis (the fourth central moment of the PDF defined in
 eq. (\ref{kurtosis}):a and b)
 and the skewness (the third central moment of the PDF defined in
 eq. (\ref{skewness}):c and d)
 for the three components of velocity and magnetic field are shown. The
 black, blue and red lines show the results in cases M0, M1, and M2.
 \label{sk}}
\end{figure}

Fig. \ref{pdf} shows the probability density functions (PDFs) for the
three components of velocity
($v_r$, $v_\theta$, and $v_\phi$) and the three components of the magnetic
field ($B_r$, $B_\theta$, and $B_\phi$), the radial vorticity
$\omega_r$, the horizontal divergence $\zeta$ and the temperature
perturbation $T_1$ in case M0 at $t=115\ \mathrm{day}$. Although the
rotation is not included in this
study, some features of the PDFs are
similar to findings from previous studies including rotation
\citep{2004ApJ...614.1073B,2008ApJ...673..557M}.
In this study, the PDF is the normalized histogram on a
horizontal surface, corrected for the grid convergence at the
poles. 
Fig. \ref{pdf}a shows the
significant asymmetry in the radial velocity. This reflects the asymmetry
between up- and downflows due to stratification (see
also Figs. \ref{area} and \ref{aflux}). The horizontal velocities
($v_\theta$ and $v_\phi$) almost show a Gaussian distribution (see
also Fig. \ref{sk}). The magnetic fields have high intermittency compared
with the velocities \citep{1996JFM...306..325B}.
Despite our asymmetric initial condition
for the longitudinal magnetic field $B_\phi=100\ \mathrm{G}$, the PDFs
show a close to symmetric distribution peaked at zero 
after a sufficiently long temporal
evolution.
The maximum value of the strength of the magnetic field is
around $10^4\ \mathrm{G}$.
The PDF for the radial vorticity is similar to that of the magnetic field,
i.e., with high intermittency. This can be expected based on the similarity between the 
induction and vorticity equations.
 The horizontal divergence has
similarity to the radial vorticity $\omega_r$ in the convection zone with high
intermittency, while $\zeta$ shows the asymmetry near the top boundary
is similar to the radial velocity $v_r$.
The temperature perturbation has
significant asymmetry, which also reflects the asymmetry between the
upflow and downflow. \par
In order to evaluate the influence of the location
of the boundary condition and the thickness of the cooling layer, we
investigate the moments of the PDF, in particular the
kurtosis $\mathcal{K}$ and the skewness $\mathcal{S}$ as
\begin{eqnarray}
&& \mathcal{K}=\frac{1}{\sigma^4}\int (x-\langle x\rangle)^4f(x)dx,
\label{kurtosis}
\\
&& \mathcal{S}=\frac{1}{\sigma^3}\int (x-\langle x\rangle)^3f(x)dx,
\label{skewness}
\end{eqnarray}
where $f(x)$ is the PDF, $x$ is each variable, and $\sigma$ is the
standard deviation as
\begin{eqnarray}
 && \sigma = \sqrt{\int (x-\langle x\rangle)^2f(x)dx}.
\end{eqnarray}
The kurtosis $\mathcal{K}$ and the skewness $\mathcal{S}$ denote the
intermittency and the asymmetry of the distribution, respectively.
We note that the PDF is normalized as $\int f(x)dx = 1$. 
For example the Gaussian PDF is characterized by 
$\mathcal{K}=3$ and $\mathcal{S}=0$.
Fig. \ref{sk}
shows the distribution of the kurtosis and the skewness for the velocity and
the magnetic field. As noted above, the horizontal velocities ($v_r$
and $v_\phi$) have almost a Gaussian distribution, i.e.,
$\mathcal{K}\sim3$ and $\mathcal{S}\sim0$ throughout the convection
zone. The radial velocity has high intermittency and asymmetry in the
convection zone. The magnetic field has intermittency and almost
symmetric distribution.
These features are common among cases M0, M1, and
M2. Quantitatively the kurtosis and skewness agree with each other
in all cases.
While the values in the near-surface region ($r>0.85R_\odot$) are
influenced by the two factors considered, the location of the upper boundary and
the thickness of the cooling layer, in the deeper region the values converge.
\par

\subsection{Generation and transportation of the magnetic field}\label{local_dynamo}
In this subsection we investigate the generation and transport of
the magnetic field by turbulent thermal convection.
The global structure of the
mutual interaction between the plasma and the magnetic field is our paramount interest.
\par
\begin{figure}[htbp]
 \centering
 \includegraphics[width=16cm]{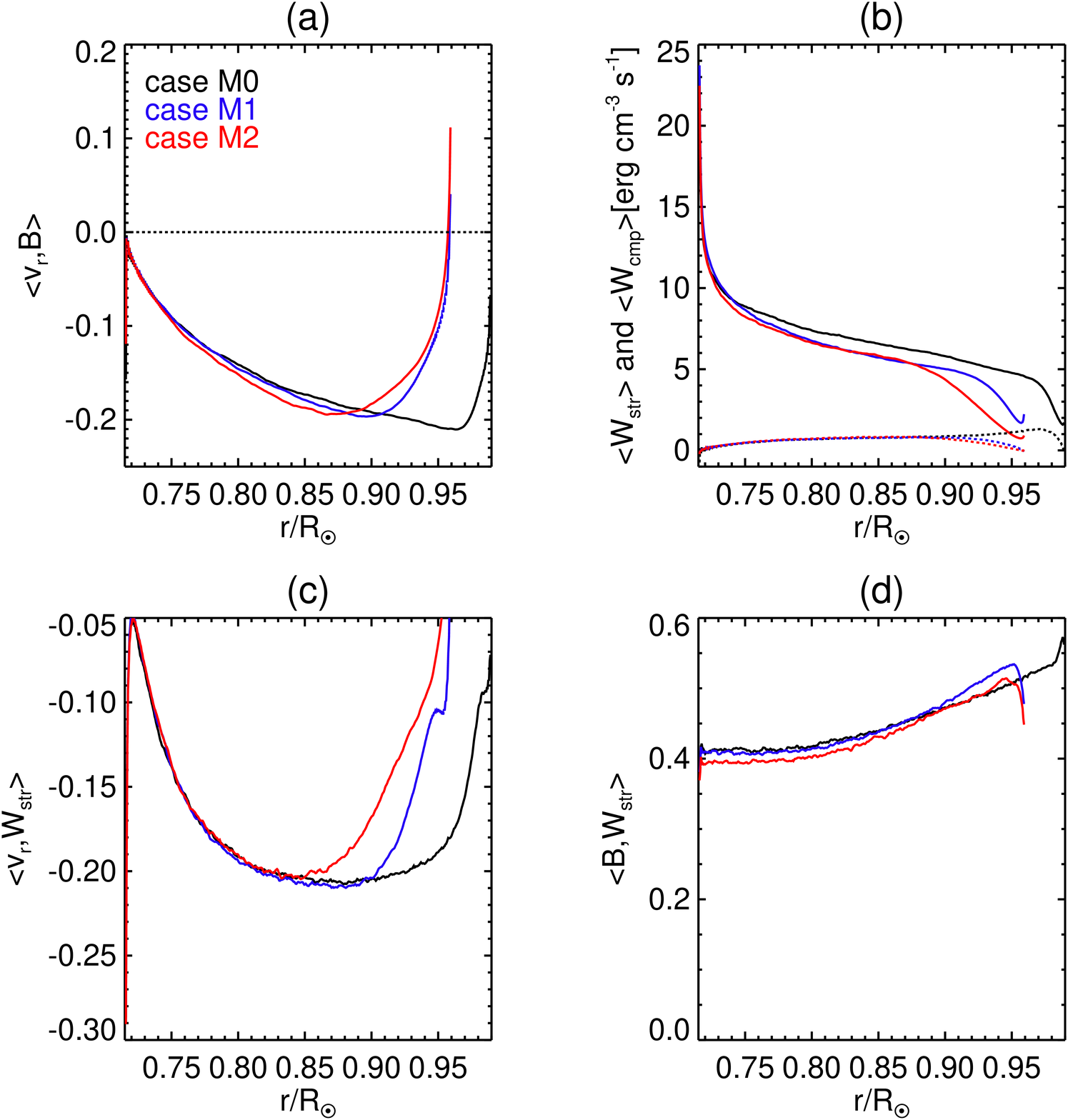}
 \caption{ 
 (a) The correlation between the radial velocity and the strength of the
 magnetic field ($\langle v_r,B\rangle$), (b) the horizontal average of the
 generation rate of the magnetic energy by the stretching 
($\langle  W_\mathrm{str}\rangle$: solid line) and the compression
($\langle W_\mathrm{cmp} \rangle$: dashed line), (c) the correlation between the
 radial velocity and the generation rate of the magnetic energy 
($\langle v_r,W_\mathrm{str}\rangle$), (d) the correlation between the
 strength of the magnetic field and the generation rate of the magnetic
 energy ($\langle B,W_\mathrm{str}\rangle$) are shown.
 \label{mag_gen}}
\end{figure}
\begin{figure}[htbp]
 \centering
 \includegraphics[width=16cm]{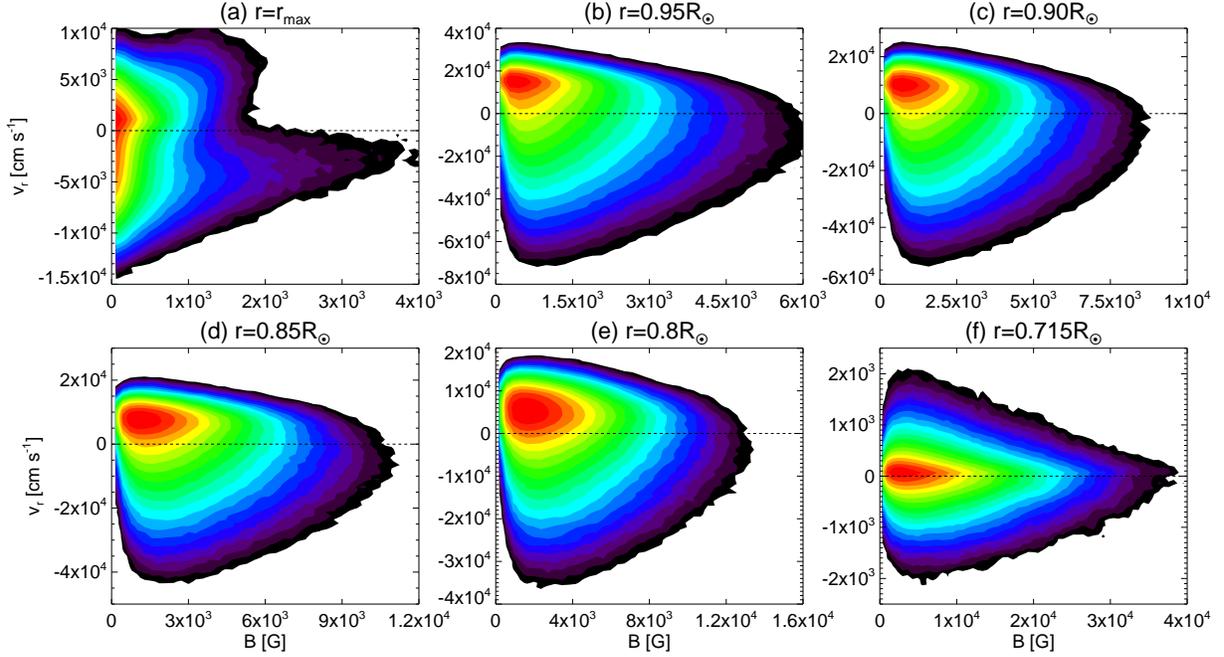}
 \caption{ Joint PDFs with the radial velocity ($v_r$) and the strength
 of the magnetic field ($B$) in the case M0 at
 (a) $r=r_\mathrm{max}$,
 (b) $r=0.95R_\odot$, (c) $r=0.90R_\odot$, (d) $r=0.85R_\odot$, 
 (e) $r=0.80R_\odot$, (f) $r=r_\mathrm{min}=0.715R_\odot$.
 The black lines show $v_r=0$, which distinguish the upflow and the
 downflow. 
 \label{pdf2d}}
\end{figure}

Fig. \ref{H0_multi} shows the radial velocity $v_r$, the radial
magnetic field $B_r$, and the radial vorticity $\omega_r$ at 
different depths ($r=0.99R_\odot$:a, b, and c, $r=0.95R_\odot$:d, e,
and f, $r=0.85R_\odot$:g, h, and i). As introduced in \S
\ref{structure}, there is good coincidence between the downflow and 
regions with a large amplitude of the radial vorticity. This means that
downflows, especially in the deeper region, include most of the turbulent
small-scale horizontal motions. 
We can also see the preferential association of a strong magnetic field with
downflows and regions with strong vorticity in the constant-depth plane (middle and right columns of
Fig. \ref{H0_multi}) and also in the meridional plane (Fig. \ref{H0_meri}).
In order to investigate this aspect quantitatively,
we estimate the correlation between the radial velocity $v_r$ and the
absolute value of the magnetic field $B$, i.e., $\langle v_r, B\rangle$ in
Fig \ref{mag_gen}a where our definition of the correlation between quantity $A$
and $B$ is defined as:
\begin{eqnarray}
 \langle A,B\rangle = \frac{\int AB dS}{\sqrt{\int A^2dS}\sqrt{\int B^2 dS}}.
\end{eqnarray}
Fig. \ref{mag_gen}a shows the negative value in most of the
convection zone. 
This means that the magnetic field
is preferentially found in downflows.
This is also seen in the joint PDFs of $v_r$ and $B$ in
Fig. \ref{pdf2d}. The figure shows the asymmetric distribution about the
$v_r=0$ axis in the convection zone. A strong magnetic field is more likely
to be located in downflow regions. We note that a symmetric
distribution is found at $r=r_\mathrm{min}$, which corresponds to
$\langle v_r, B\rangle\sim0$ in Fig. \ref{mag_gen}a.\par
There are two possibilities for the preference of the magnetic field to
downflow regions. One is
that the
magnetic field is generated uniformly in space and transported to the
downflow region by converging motion. The other is that the magnetic
field is generated in the downflow region. In order to answer
this question, we define and evaluate the generation rate of
magnetic energy by
the stretching ($W_\mathrm{str}$) and the compression ($W_\mathrm{cmp}$)
as:
\begin{eqnarray}
 && W_\mathrm{str} = \frac{\bf B}{4\pi}\cdot \left[ (\bf B\cdot \nabla)
					      \bf v\right],\\
 && W_\mathrm{cmp} = -\frac{B^2}{4\pi}(\nabla\cdot {\bf v}).
\end{eqnarray}
The estimations for the horizontal averages,
$\langle W_\mathrm{str}\rangle$ and $\langle W_\mathrm{cmp} \rangle$, are
shown in Fig \ref{mag_gen}b. It is clear that the value of the
stretching $\langle W_\mathrm{str}\rangle$ is much larger than that of the
compression throughout the convection zone and that the generation of the magnetic
field is basically done by
the stretching in the turbulent motion. This is also seen by the
realistic calculations in the photosphere \citep{0004-637X-714-2-1606}, which  suggest that 95\% of the gain of the magnetic energy is done by
the stretching.
We also estimate the correlation
between the radial velocity $v_r$ and the energy generation rate by the
stretching $W_\mathrm{str}$ (Fig. \ref{mag_gen}c). The distribution of
this correlation is similar to that of $\langle v_r, B\rangle$, i.e., the
effective stretching prefers the downflow region. Thus we conclude
that the reason the strong magnetic field prefers the downflow
region is that the magnetic field is more likely to be generated there.
The correlation between $B$ and $W_\mathrm{str}$ is also
estimated. This shows a larger value of ($>0.4$) throughout the convection zone
(Fig. \ref{mag_gen}d).
The features discussed above are basically common among the studied
cases.\par
\begin{figure}[htbp]
 \centering
 \includegraphics[width=14cm]{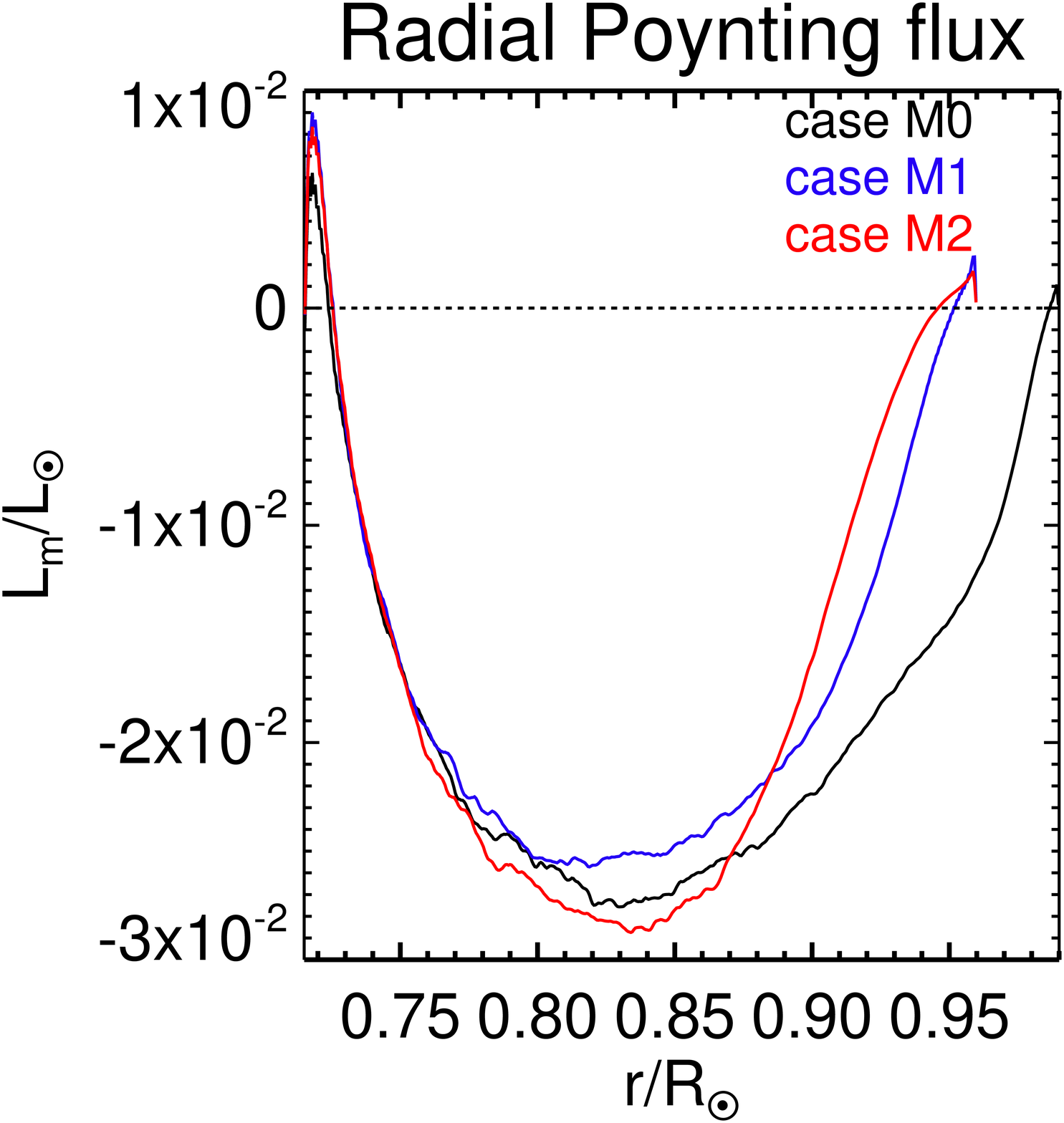}
 \caption{ The integrated radial Poynting flux as a function of the depth is
 shown. The black, blue and red lines show the results
 in cases M0, M1, and M2, respectively.
 \label{poynting_flux}}
\end{figure}

In order to investigate magnetic field transport in the convection
zone, we evaluate the Poynting
flux given by:
\begin{eqnarray}
 F_\mathrm{m} = \frac{c}{4\pi}
  \left(E_\theta B_\phi - E_\phi B_\theta\right),
\end{eqnarray}
where the electric field is defined as
${\bf E}=-({\bf v}\times{\bf B})/c$ and $c$ is the speed of light (see
\cite{2004ApJ...614.1073B}). Fig. \ref{poynting_flux} shows the
horizontally integrated Poynting flux $L_\mathrm{m}$ as a function of the
depth. Since the
absolute value ($\sim 10^{31}\ \mathrm{erg\ s^{-1}}$) is much smaller than
the solar luminosity ($L_\odot=3.84\times10^{33}\ \mathrm{erg\
s^{-1}}$), the Poynting flux does not contribute significantly to the total energy flux
balance. The Poynting flux is negative in most of the
convection zone, since a strong magnetic field is concentrated in
downflow regions. This is suggested by previous studies as the
turbulent pumping effect
\citep[see, e.g.,][]{1998ApJ...502L.177T,2001ApJ...549.1183T}. Within the convection
zone, the flux has a positive value only in the thin layer ($\sim
0.01R_\odot$) close to the bottom. This is caused by the bounced motion
from the bottom
boundary. The magnetic energy is transported upward, causing the
magnetic flux in the upflow region. This can be one of the reasons why
the absolute values of the correlation $\langle v_r,B\rangle$ is small
in the deeper region (Fig. \ref{mag_gen}a) and why the
distribution of the joint PDF in the bottom (Fig. \ref{pdf2d}f) is symmetric.
\par
\begin{figure}[htbp]
 \centering
 \includegraphics[width=16cm]{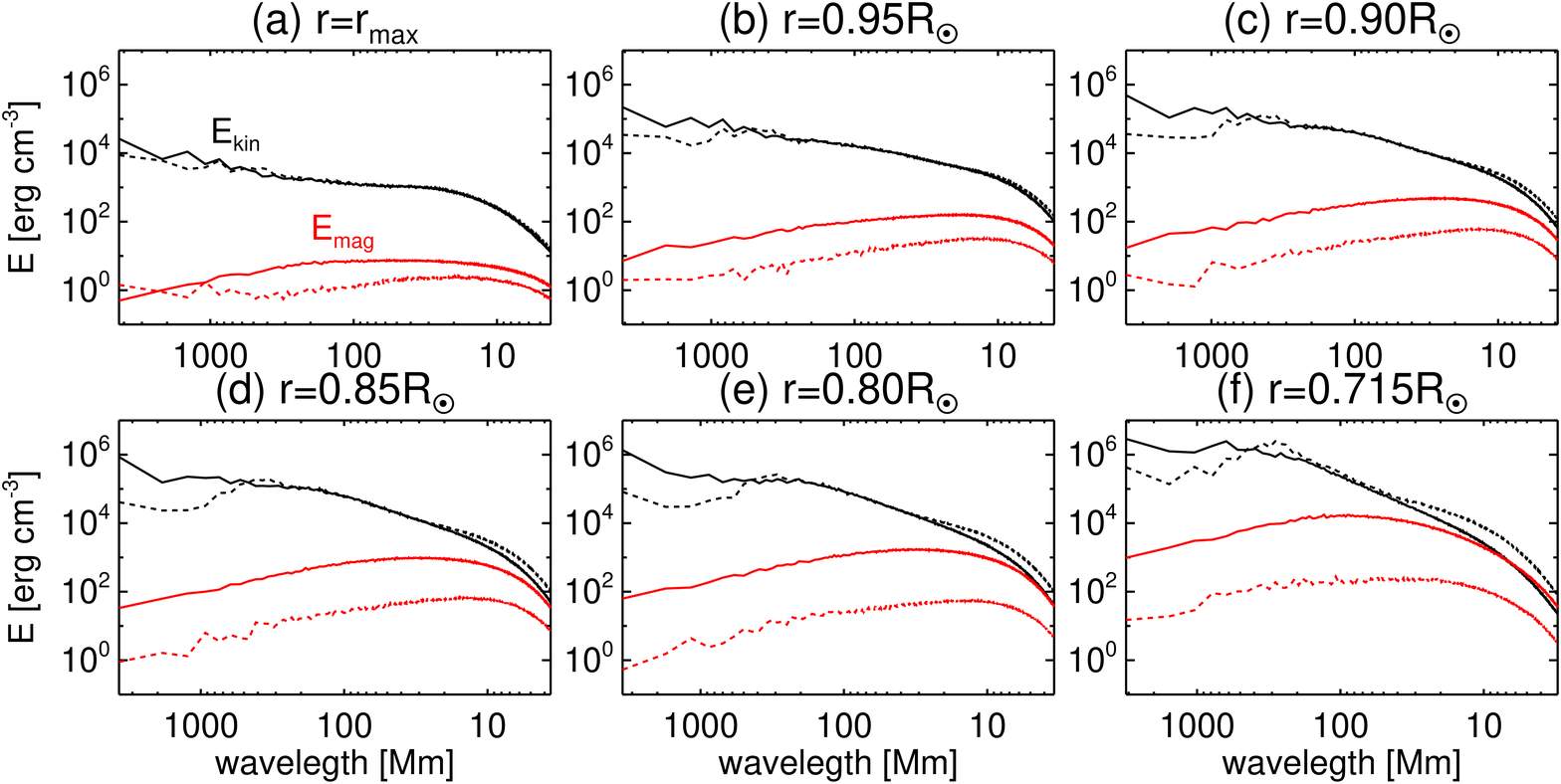}
 \caption{ The spectra of the kinetic energy
 ($E_\mathrm{kin}=\rho_0v^2/2$: solid black lines) and the magnetic energy 
($E_\mathrm{mag}=B^2/(8\pi)$: solid red lines) in case M0. The
 dashed black lines show the
 kinetic energy in H0, i.e., without the magnetic field. The dashed red
 lines show the magnetic energy at $t=5.8$ days.
 \label{magkin}}
\end{figure}

In the following discussion, we investigate the scale of the magnetic field
generated.
Fig. \ref{magkin} shows the spectra of the kinetic energy
(black lines: $E_\mathrm{kin}=\rho_0v^2/2$)
and the magnetic energy (red lines: $E_\mathrm{mag}=B^2/(8\pi)$)
in case M0 averaged from $t=173\ \mathrm{day}$ to 
$t=237\ \mathrm{day}$ in which the generation of the
magnetic field is close to being saturated. 
The dashed black and red lines show
the kinetic energy without the magnetic field and the magnetic energy at
$t=5.8$ days, respectively.
In the upper convection zone ($\geq0.85R_\odot$), the spectra of the
magnetic energy peaks at the smallest scale, which is typical for
the kinematic phase of a local dynamo. Finding this feature in the saturated
phase indicates that the local
dynamo is likely not very efficient for our resolved scales ($>7\ \mathrm{Mm}$).
We see also no indication that the kinetic energy spectrum changed due to the presence of the dynamo near
the top of the domain. This situation is different in the lower half of the domain.
Figs. \ref{magkin}e and f
show some peak shift of the magnetic energy to larger scales and some
feedback on thermal convection, i.e.,  the kinetic
energy is suppressed on the smallest scales. Here the
magnetic energy slightly exceeds the kinetic energy near the smallest
scales. This shows that a more efficient local dynamo can be achieved to
some extent
for our resolution in the lower part of the convection zone ($<0.85R_\odot$).
\par
\begin{figure}[htbp]
 \centering
 \includegraphics[width=16cm]{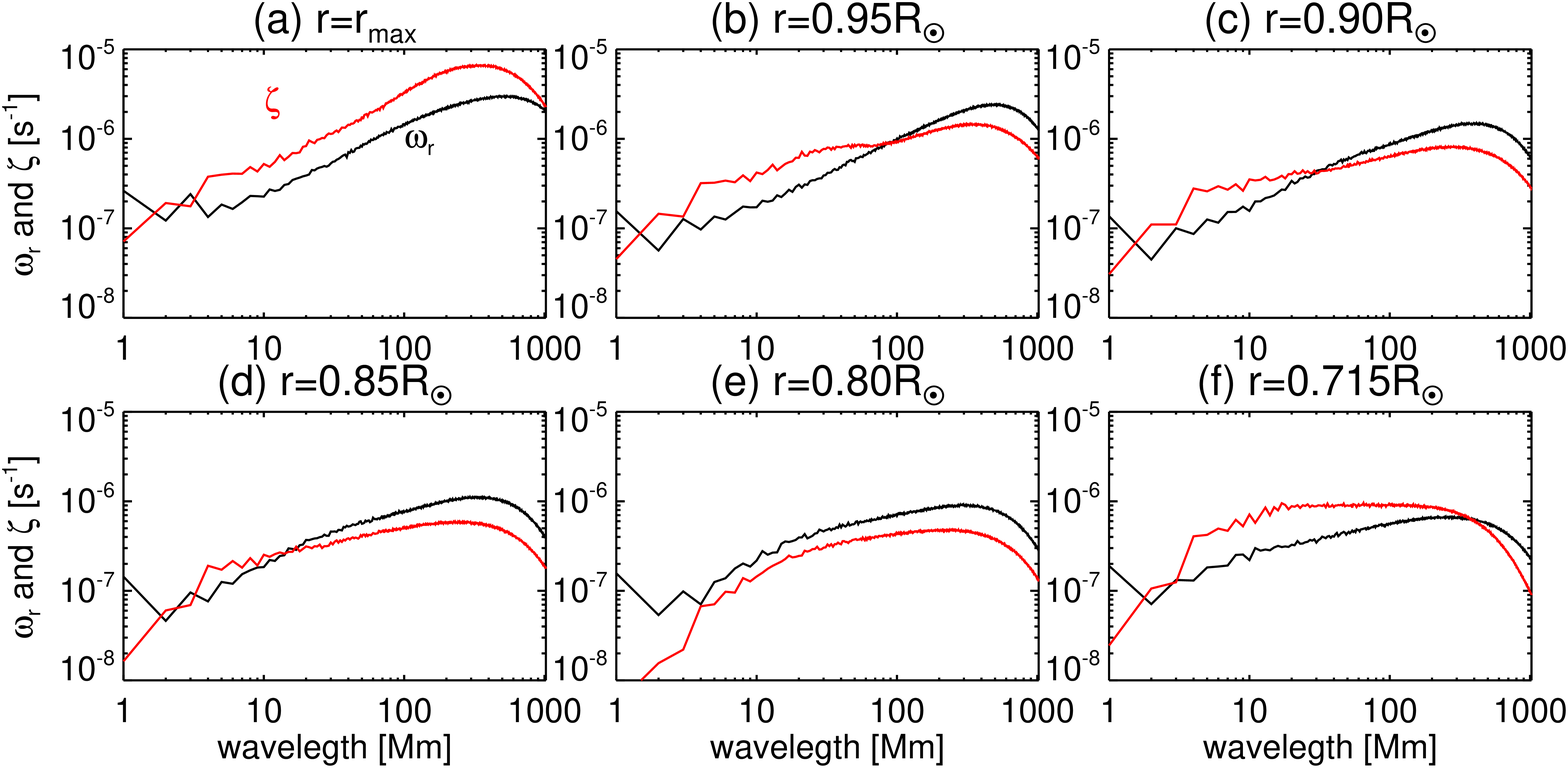}
 \caption{ The spectra of the horizontal divergence $\zeta$ and the
 radial vorticity $\omega_r$ using the result in case M0.
 \label{omdi}}
\end{figure}

Fig. \ref{omdi} shows that the spectra of the horizontal divergence ($\zeta$)
and the radial vorticity ($\omega_r$) show a similar
distribution to that of the magnetic field with the peak at small
scales. \cite{1961JFM....11..625M} suggested that the power spectrum
varies as $D_b(k)\propto k^2D_v(k)$ when the magnetic field is
proportional to $(\nabla\cdot {\bf v})$ or $(\nabla\times{\bf v})$,
where $D_b(k)$ and $D_v(k)$ are magnetic and kinetic spectra. 
This was recently confirmed through solar observations in the photosphere using the
{\it Hinode}
satellite \citep{2012ApJ...758..139K}. Since in our calculation there is
similarity between the magnetic field and the vorticity or the
divergence, the magnetic energy peaks at smaller scales
than does the kinetic energy. We note that differences can be seen between
the shear of the velocity ($\omega_r$ and $\zeta$) and the magnetic energy
($E_\mathrm{mag}$) at the base of the convection zone (Fig. \ref{magkin}f and
\ref{omdi}f), where the feedback from the magnetic field is stronger.
\par
Even though the numerical resolution 
does not vary much with depth, the effectiveness of the local dynamo is not expected to be depth 
independent. There are two reasons for this: The intrinsic convective scale varies with depth, and
a downward Poynting flux exists almost everywhere in the convection
zone. 
In the upper half of the
convection zone the divergence of the Poynting flux provides an energy sink,
while at the same time the dynamo is not very efficient on the 
resolved scale
\citep[similar discussion can be found in][]{2003ASPC..286..121S}. 
In the lower convection zone ($< 0.85R_\odot$), the radial gradient of
the Poynting flux is negative, thus the magnetic energy is accumulated.
At the same time the intrinsic scale of convection is larger and
better resolved, leading to a more efficient dynamo.
It should be noted that \cite{2007A&A...465L..43V} found that even in near-surface areas the small-scale convection reproduced with high resolution
has a time scale short enough to amplify
the magnetic field against the pumping effect. 
\par
\begin{figure}[htbp]
 \centering
 \includegraphics[width=12cm]{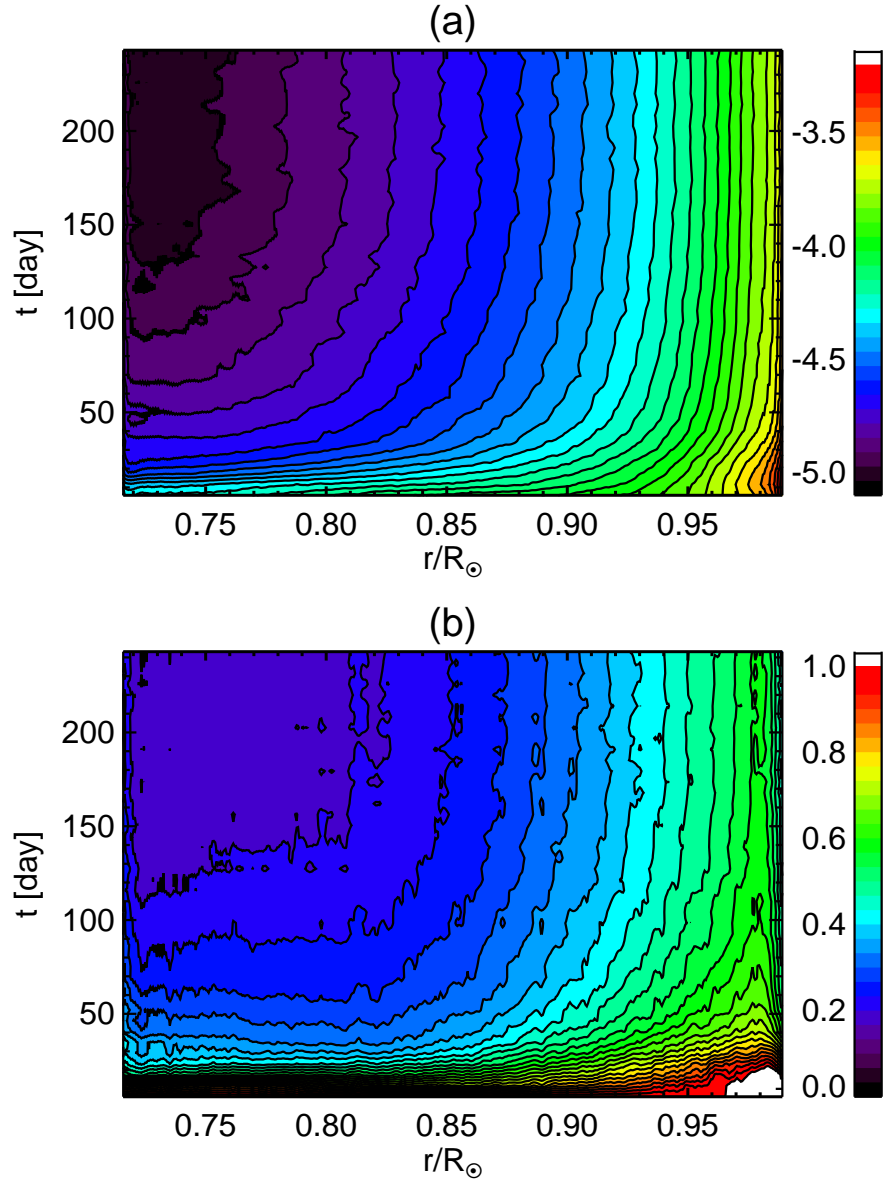}
 \caption{ The time evolution of the effective shear (a)
$\log f_\mathrm{eff}$ and (b) $f_\mathrm{eff}(t)/f_\mathrm{eff}(t=5.8\ \mathrm{day})$.
 \label{wstr_time}}
\end{figure}

In order to confirm our idea about the generation and transportation of
the magnetic field in our calculation, we evaluate the effective shear for the
magnetic field,
\begin{eqnarray}
 f_\mathrm{eff} = \frac{\langle W_\mathrm{str}\rangle}{\langle B^2/(8\pi)\rangle}.
\end{eqnarray}
$f_\mathrm{eff}$ has the unit of $\mathrm{s^{-1}}$ and indicates the time scale of
the amplification. Fig. \ref{wstr_time}a shows the time evolution of
$\log f_\mathrm{eff}$ and indicates that regardless of the phase of the
generation of the magnetic field, the larger value of $f_\mathrm{eff}$ is seen in the upper
convection, which reflects the short time scale of the thermal
convection there. At later times with a stronger magnetic field, the effective shear $f_\mathrm{eff}$
is reduced, and after $t\sim150$, it remains
constant. Fig. \ref{wstr_time}b shows the ratio to the value at 
$t=5.8\ \mathrm{day}$, i.e. 
$f_\mathrm{eff}(t)/f_\mathrm{eff}(t=5.8\ \mathrm{day})$.
The suppression of the effective shear depends on the depth.
It is more suppressed in the deeper region. This might be caused by the
feedback of the magnetic field on the velocity seen in Fig. \ref{magkin}
(the solid and dashed black lines) as well as a misalignment between the shear
and magnetic field around the base of the convection zone. The ineffective
suppression of $f_\mathrm{eff}$ in the upper convection zone confirms
our idea presented in the previous paragraph. The local dynamo saturates there
with little non-linear feedback (suppression of $f_\mathrm{eff}$), since the pumping
effect works well in the upper region and the dynamo is not very efficient on the
resolved scales to begin with.
\par
\begin{figure}[htbp]
 \centering
 \includegraphics[width=12cm]{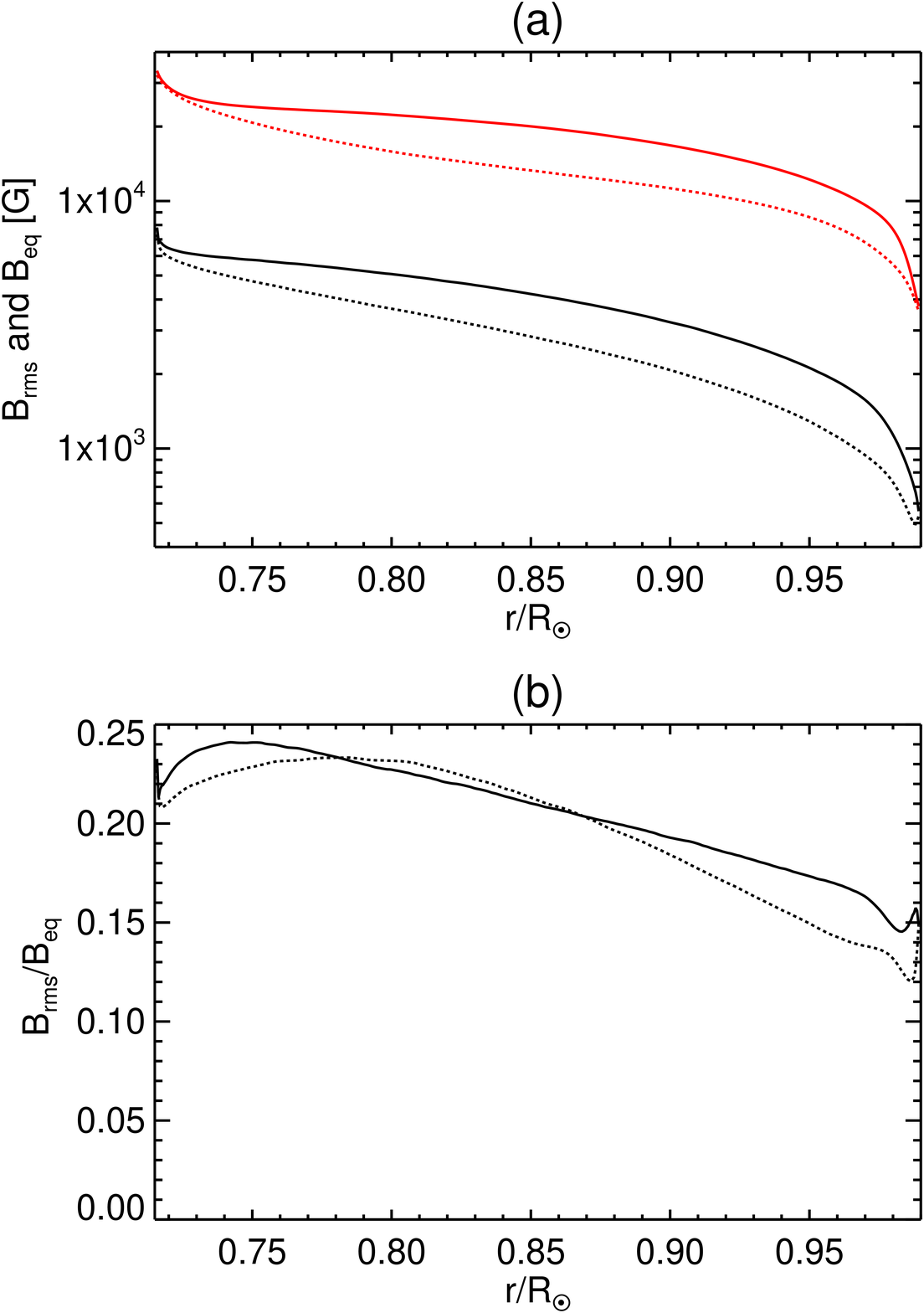}
 \caption{ (a) The distribution of the equipartition field
 $B_\mathrm{eq}$ (red lines) and the RMS magnetic field $B_\mathrm{rms}$
 (black lines).
 (b) The ratio of the RMS magnetic field and the equipartition magnetic
 field ($B_\mathrm{rms}/B_\mathrm{eq}$). The solid and dashed lines show the
 values at the downflow and the upflow regions, respectively.
 \label{brmseq}}
\end{figure}

Fig. \ref{brmseq} also shows
the variation of dynamo efficiency
in the convection zone.
Fig. \ref{brmseq}a shows the
equipartition magnetic field 
$B_\mathrm{eq} = \sqrt{4\pi v_\mathrm{rms}^2}$ and the RMS magnetic
field $B_\mathrm{rms}$ as functions of the depth. The solid and dotted
lines show the values at
the downflow and upflow regions, respectively.  Basically, both
$B_\mathrm{eq}$ and $B_\mathrm{rms}$ increase with the depth
and, Fig. \ref{brmseq}b shows an increase of $B_\mathrm{rms}/B_\mathrm{eq}$
with depth, which is caused by the ineffectiveness of the local dynamo in
the upper region due to the pumping effect and our insufficient resolution.
\par
\section{Discussion and Summary}\label{discussion}
We carry out the high-resolution calculations of the solar global
convection which resolve  10 Mm-scale
convection smaller than the supergranulation using the RSST (described in \S \ref{S:rsst}).
The RSST leads to a simple algorithm and requires only local communication in
the parallel computing. In addition, this method has the capability to access
the real solar surface without undermining the important physics.
This enables us to capture near-surface small-scale convection while
keeping a global domain.
Our main conclusions are given as follows: 1. Small-scale convection
is excited close to the surface ($>0.9R_\odot$), when we
expand our domain upward to $0.99\,R_{\odot}$ to capture
the near-surface layers with small pressure scale heights. 2. In deeper convection
zones ($<0.9R_\odot$), the convection flow is not influenced by the location of the top boundary and the
assumed thickness of the thermal boundary layer. 
Changing the position of the top boundary or thickness of the cooling layer does
not lead to significant differences
in the
convective structure and properties of the local dynamo, in terms of the
power-spectrum, the probability density function and the local dynamo.
3. Using an LES approach we can achieve small-scale dynamo action and
maintain a field of about $0.15-0.25 B_\mathrm{eq}$ throughout the convection
zone.
4. The overall dynamo efficiency varies significantly in the convection
zone as a consequence of the downward directed Poynting flux and the
depth variation of the intrinsic convective scales. For a fixed numerical resolution,
the dynamo-relevant scales are better resolved in the deeper convection zone and
are therefore less affected by numerical diffusivity, i.e., the effective Reynolds numbers
are larger.
\par
Recently, \cite{2012PNAS..10911928H} suggested that there is a
significant difference between the calculation by
\cite{2008ApJ...673..557M} and flow velocities inferred
from local
helioseismology. The observed convection amplitude is much smaller than that in the
hydrodynamic calculation. We expected that the modification to the
location of the upper boundary may help to
resolve this issue.
Conclusion 1 above, however, suggests that the difference between the
calculation and
the observation becomes larger when we
move the top boundary further upward. We see, nevertheless, some indication
that the kinetic energy is reduced on the largest scales when a local dynamo is
present.
At the same time,
we cannot fully rule out the
possibility that unresolved small-scale turbulence near the top boundary
contributes to this issue,
since the intrinsic scale of the convection near our top boundary is
still not well resolved.
\par
Conclusion 2 is one of the most important results in this study,
since it suggests that previous calculations such as by
\cite{2008ApJ...673..557M} are physically reasonable in the
deeper convection zone even if the top boundary
condition is placed significantly below the solar surface.
\par
\begin{figure}[htbp]
 \centering
 \includegraphics[width=16cm]{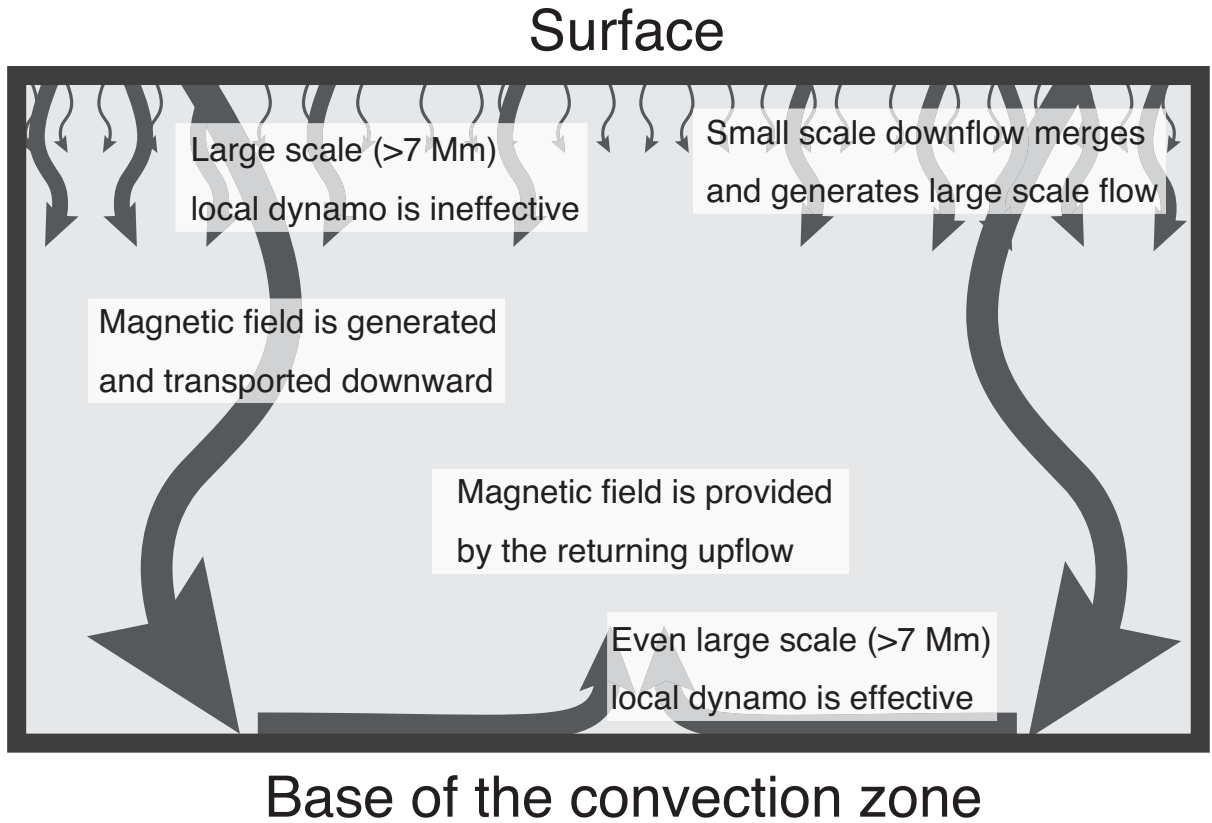}
 \caption{
The schematic view of our calculation.
 The arrows show the convection flow.
The conclusion that the
dynamo is ineffective near the top is based on the comparison of the
kinematic and
saturated spectra and the only moderate reduction of the effective
shear in the upper convection zone (see \S \ref{local_dynamo}).
 \label{local_dynamo_fig}}
\end{figure}

We summarize our simulation results in a schematic picture shown in
Fig. \ref{local_dynamo_fig}.
The conclusion that the
dynamo is ineffective near the top comes from the comparison of the
kinematic and
saturated spectra and the only moderate reduction of the effective
shear in the upper convection zone (see \S \ref{local_dynamo}). 
Since several aspects, in particular with regard to the local dynamo,
require higher
resolution before being directly applicable to the solar convection zone,
Fig. \ref{local_dynamo_fig} does not necessarily apply to the real Sun.
High resolution simulations of a local dynamo in the solar photosphere
\citep{2007A&A...465L..43V} suggest that 
efficient dynamo action
is possible even in the presence of the pumping effect. 
These simulations use, however,
a grid resolution of about a factor of $100$ larger than our setup, which is currently
inapplicable for global scale convection simulations. Therefore, the dynamo RMS field strength of
$0.15-0.25 B_\mathrm{eq}$ can likely be considered a lower limit.
 \par
One issue we cannot address in this study is
the problem regarding the small magnetic Prandtl number ($\mathrm{Pm}\sim
10^{-3}$) in the solar
convection zone, since we adopt numerical diffusivities which assume that
the magnetic Prandtl number is around unity.
Several authors argued that smaller magnetic Prandtl numbers
make local dynamos less efficient
\citep[e.g.][]{2004PhRvL..92e4502S,2004PhRvL..92n4501B}.
In other words, a small magnetic Prandtl number requires a larger
magnetic Reynolds number 
in order to achieve a super-critical dynamo.
While this can be
a significant problem for numerical simulations with rather moderate
Reynolds numbers, this is less likely an issue in the solar convection zone
with Reynolds numbers as large as $\mathrm{Rm}\sim 10^{11}$ 
\citep{2011ApJ...741...92B}.
Thus we believe
that our approach relying only on numerical diffusivities can capture the
physics of the local dynamo in the solar convection zone.
\par
Rotation is not included in this study although it plays a key role in organizing
large-scale flows and enabling large-scale dynamo action.
Studying the near-surface small-scale convection in a setup with rotation is
of particular interest to us for investigation of the origin of the near-surface shear layer
observed on the Sun. Our investigation of this is a work-in-progress and will be presented
in a forthcoming paper. Further progress can be made by using non-uniform
grids, such as nested grids or adaptive mesh refinement, which are useful
in order to overcome the significant differences in spatial and temporal dynamic scales.
These methods are already implemented in our numerical code
\citep{2012A&A...548A..74H}.

\acknowledgements
The authors thank M. Miesch for helpful comments on the draft.
H. H. is supported by Grant-in-Aid for JSPS Fellows.
The National Center for Atmospheric Research is sponsored by the
National Science Foundation.
Part of the results was obtained by using the K computer at the RIKEN
Advanced Institute for Computational Science (Proposal number hp120287).
This research was partly conducted using the Fujitsu PRIMEHPC FX10 System
(Oakleaf-FX) in the Information Technology Center, The University of
Tokyo.
The authors are grateful to Ryoji Matsumoto for managing our
computational resources.
We have greatly benefited from the proofreading/editing assistance
from the GCOE program.
This work is partially supported by ``Joint Usage/Research Center for
Interdisciplinary Large-scale Information Infrastructures'' 
 ``High Performance Computing Infrastructure''
in Japan.

\clearpage
\appendix
\section{Equation of state in the solar convection zone}
\label{app:eos}
In the numerical simulations of the convection in the near-surface layer,
the ordinary tabular equation of state is widely used
\citep{2005A&A...429..335V,2009ApJ...691..640R}.
However, this is not a viable approach for our current simulations,
since the deviations from the reference state
are small, e.g., $\rho_1/\rho_0 \sim 10^{-6}$ around the base of
the convection zone. Thus we adopt another way to treat the ionization effect
in the near-surface layer. The fluctuations from the reference state are
calculated as:
\begin{eqnarray}
&& p_1 = \left(\frac{\partial p}{\partial \rho}\right)_s   \rho_1 
       + \left(\frac{\partial p}{\partial s   }\right)_\rho s_1,\\
&& T_1 = \left(\frac{\partial T}{\partial \rho}\right)_p   \rho_1 
       + \left(\frac{\partial T}{\partial p   }\right)_\rho p_1,\\
&& e_1 = \left(\frac{\partial e}{\partial \rho}\right)_T   \rho_1 
       + \left(\frac{\partial e}{\partial T   }\right)_\rho T_1.
\end{eqnarray}
The first derivatives, such as $(\partial p/\partial \rho)_s$, are
described by the background variables, $\rho_0(r)$, $p_0(r)$... and are
regarded as functions of the depth $r$. In the OPAL
routine \citep{1996ApJ...456..902R}, the values 
$(\partial e/\partial \rho)_T$,
$(\partial e/\partial T)_\rho$, $(\partial p/\log \rho)_T$
and $(\partial p/\log T)_\rho$ are provided for given $\rho_0$,
$T_0$, and 
the mass fraction of hydrogen ($X$), helium ($Y$) and other
metals ($Z$). 
We show the relations between the OPAL-provided variable and the
required variable derived from the first law of thermodynamics
\citep{1984oup..book.....M}.
\begin{eqnarray}
&& \left(\frac{\partial p}{\partial \rho}\right)_s =
   \frac{c_\mathrm{p}}{\kappa_\mathrm{t}\rho_0 c_\mathrm{v}},\\
&& \left(\frac{\partial p}{\partial s}\right)_\rho =
 \frac{\beta_\mathrm{p}T_0}{\kappa_\mathrm{t}c_\mathrm{v}},\\
&& \left(\frac{\partial T}{\partial \rho}\right)_p = -\frac{1}{\rho_0
 \beta_\mathrm{p}},\\
&& \left(\frac{\partial T}{\partial p}\right)_\rho = T_0/
\left(\frac{\partial p}{\partial \log T}\right)_\rho,
\end{eqnarray}
where $\beta_p$, $c_\mathrm{v}$, $c_\mathrm{p}$ and $\kappa_\mathrm{t}$
are the
coefficient of thermal expansion, the specific heat at constant volume
and pressure, the coefficient of isothermal compressibility, respectively,
defined as:
\begin{eqnarray}
&& \beta_\mathrm{p} \equiv -\left(\frac{\partial \log \rho}{\partial T}\right)_p
  = \frac{1}{T_0}\left(\frac{\partial p}{\partial \log T}\right)_\rho/
  \left(\frac{\partial p}{\partial \log \rho}\right)_T,\\
&& c_\mathrm{v} = \left(\frac{\partial e}{\partial T}\right)_\rho,\\
&& c_\mathrm{p} = c_\mathrm{v}
-T_0\beta_\mathrm{p}\left[\left(\frac{\partial e}{\partial \rho}\right)_T
       -\left(\frac{p}{\rho^2}\right)\right],\\
&&\kappa_\mathrm{t} = \left(\frac{\partial \log \rho}{\partial p}\right)_T.
\end{eqnarray}

\section{Conservation of the total energy in the RSST and first-order enthalpy flux}
\label{enthalpy_flux}
When the equation of the entropy is given as eq. (\ref{eqen}), the total energy is
not conserved even mathematically.
In this appendix, the deviation in the conservation of the total energy
caused by the RSST is introduced. 
Here , we ignore the magnetic field, the gravity and the
radiation for the sake of simplicity.
The equation of continuity for RSST is defined as eq. (\ref{eqco}).
Thus there are two relations as:
\begin{eqnarray}
 \frac{D\rho}{Dt}
&=&\frac{\partial \rho}{\partial t} + {\bf v}\cdot \nabla
  \rho\nonumber\\
&=&-\frac{1}{\xi^2}\nabla\cdot(\rho {\bf v})
 + {\bf v}\cdot \nabla \rho\nonumber\\
&=& \left(1-\frac{1}{\xi^2}\right)\nabla\cdot(\rho {\bf v})-\rho\nabla\cdot {\bf v},
\end{eqnarray}
and
\begin{eqnarray}
 \rho\frac{D A}{Dt} &=& \rho\frac{\partial A}{\partial t} + \rho{\bf
  v}\cdot \nabla A\nonumber\\
 &=&\frac{\partial}{\partial t}(\rho A) + \nabla\cdot\left(\rho A {\bf
						      v}\right)
 -A\frac{\partial \rho}{\partial t} - A\nabla\cdot(\rho {\bf v})\nonumber\\
 &=&\frac{\partial}{\partial t}(\rho A) + \nabla\cdot\left(\rho A {\bf
						      v}\right)
-A\left(1-\frac{1}{\xi^2}\right)\nabla\cdot(\rho {\bf v}).
\end{eqnarray}
There is deviation from the relation with the original equation of
continuity. The deviations are proportional to
$(1-1/\xi^2)\nabla\cdot(\rho{\bf v})$.
Therefore the deviation between the conservative form and the primitive
form of the equation of motion is
\begin{eqnarray}
 \left(1-\frac{1}{\xi^2}\right){\bf v}\nabla\cdot\left(\rho {\bf v}\right).
\end{eqnarray}
Then the deviation between the equation of entropy and the conservative
form for the total energy is introduced as:
\begin{eqnarray}
 \rho T\frac{Ds}{Dt}&=&
\rho\frac{De}{Dt}+\rho p\frac{D}{Dt}\left(\frac{1}{\rho}\right)
			      \nonumber\\
 &=& \frac{\partial }{\partial t}\left(\rho e\right)
  + \nabla\cdot\left(\rho e
							  {\bf v}\right)
 + p\nabla\cdot{\bf v}\nonumber \\
 &&- \left(e+\frac{p}{\rho}\right)\left(1-\frac{1}{\xi^2}\right)\nabla\cdot(\rho {\bf v})
 \nonumber\\
 &=& Q,
\end{eqnarray}
where $Q$ includes thermal diffusivity and surface cooling.
When we start from the primitive form of the equation of motion, the
equation of the kinetic energy is expressed as:
\begin{eqnarray}
 \frac{\partial }{\partial t}\left(\frac{1}{2}\rho v^2\right)
  + \nabla\cdot\left(\frac{1}{2}\rho v^2{\bf v}\right) =
  -{\bf v}\cdot\nabla p  +\frac{1}{2}
  v^2\left(1-\frac{1}{\xi^2}\right)\nabla\cdot(\rho {\bf v}).
\end{eqnarray}
Thus the conservative form for the total energy is expressed as:
\begin{eqnarray}
 \frac{\partial}{\partial t}\left(\rho e +\frac{1}{2}\rho
			     v^2\right)  &=&
  -\nabla\cdot\left[\left(\rho e + p+\frac{1}{2}\rho
	      v^2\right){\bf v}\right],\nonumber\\
 && + \left( e + \frac{p}{\rho} + 
       \frac{1}{2} v^2\right)\left(1-\frac{1}{\xi^2}\right)\nabla\cdot(\rho
       {\bf v}) +Q.
       \label{total_flux}
\end{eqnarray}
If the speed of sound is much faster than the fluid velocity, the
anelastic approximation $\nabla\cdot(\rho{\bf v})=0$ is well achieved.
In our calculations over 200 days, we could not see any coherent trend in the energy.
\par
\section{First-order enthalpy flux}
We derive the first-order radial enthalpy flux. According to
eq. (\ref{total_flux}), the original form of the radial enthalpy flux is
expressed as:
\begin{eqnarray}
 F_\mathrm{e} = \left(\rho e + p\right)v_r.
\end{eqnarray}
In the statistical steady state, horizontal integration and the time
average ensure no net transport of mass along the depth, i.e.,
\begin{eqnarray}
 \int_s \rho v_rdS = 0.
\end{eqnarray}
In this study the equation of state for the perfect gas is not adopted, and
the form of the enthalpy flux is slightly different from previous studies
\cite[e.g.][]{2008ApJ...673..557M,2011A&A...531A.162K} in which the
enthalpy flux is simply expressed as $F_\mathrm{e}=c_\mathrm{p}\rho_0T_1v_r$. 
The integrated radial enthalpy flux is calculated as:
\begin{eqnarray}
 L_\mathrm{e} &=& \int_s \left(e + \frac{p}{\rho}\right)\rho
  v_r dS\nonumber\\
  &=& \int_s \left(e + \frac{p_0+p_1}{\rho_0 +
	    \rho_1}\right)\rho v_r\nonumber dS\\
  &\sim& \int_s \left[e + \frac{p_0+p_1}{\rho_0}
	       \left(1-\frac{\rho_1}{\rho_0}\right)\right]\rho
  v_r\nonumber dS \\
 &\sim& \int_s \left(e_1 +
	      \frac{p_1}{\rho_0}-\frac{p_0\rho_1}{\rho_0^2}\right)\rho
 v_r dS\nonumber\\
 &\sim& \int_s \left(\rho_0 e_1 + p_1 -
	      \frac{p_0\rho_1}{\rho_0}\right)v_r dS.
\end{eqnarray}
The final line of the equation is then adopted for the enthalpy flux in
this study.
\begin{figure}[htbp]
 \centering
 \includegraphics[width=15cm]{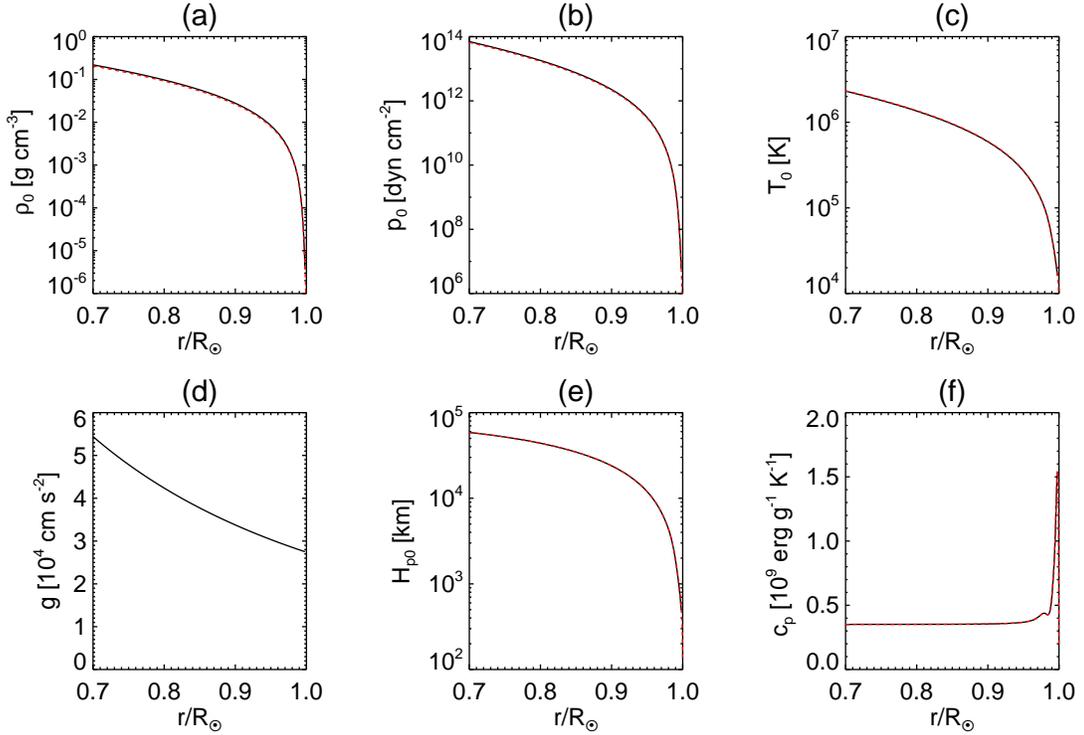}
 \caption{
The values at the reference state, (a) the density, (b) the gas pressure,
 (c) the temperature, (d) the gravitational acceleration, (e) the pressure
 scale height, (f) the heat capacity at constant pressure, are shown.
The black and red lines show the reference state in our study and the
 values from Model S, respectively. The value of the gravitational
 acceleration in this study is exactly the same as in Model S.
\label{back}}
\end{figure}

\begin{table}
\begin{center}
\caption{Important parameters in our studies.\label{param}}
\begin{tabular}{lccc}
\tableline\tableline
Case & H0 & H1 & H2 \\
\tableline
$N_r$  & 512& 456 & 456 \\
$r_\mathrm{max}/R_\odot$ & 0.99 & 0.96 & 0.96\\
$\displaystyle{\frac{\rho_0(r_\mathrm{min})}{\rho_0(r_\mathrm{max})}}$ 
& 613 & 36 & 36 \\
$H_{p0}(r_\mathrm{max})\ [\mathrm{km}]$ & 1870 & 9390 & 9390 \\
$d_\mathrm{c}\ \mathrm{[km]}$ & 3740 & 3740 & 18780 \\
\tableline
\end{tabular}
\end{center}
\end{table}

\clearpage

\clearpage

\end{document}